\documentclass[%
preprint,tightenlines,
nofootinbib,amsmath,amssymb,
aps,
floatfix
]{revtex4-1}
\pdfoutput = 1

\usepackage{graphicx}          
\usepackage{dcolumn}           
\usepackage{multirow}
\usepackage{bm}                
\usepackage{color}
\usepackage{xcolor}
\usepackage{url}

\definecolor{darkblue}{rgb}{0.0,0,0.5}
\definecolor{darkred}{rgb}{0.7,0,0.0}
\definecolor{medmagenta}{rgb}{0.3,0,0.7}
\definecolor{darkmagenta}{rgb}{0.4,0,0.4}
\usepackage[unicode=true, bookmarks=false, linkcolor = darkblue, citecolor = medmagenta, urlcolor=darkmagenta,breaklinks=false, colorlinks=true, hyperfootnotes=true]{hyperref}

\usepackage{epsfig,amssymb,amsmath,psfrag,epstopdf,color}
\usepackage{ulem}
\usepackage{amsthm}

\newcommand{\be}{\begin{equation}}
\newcommand{\ee}{\end{equation}}
\newcommand{\bea}{\begin{eqnarray}}
\newcommand{\eea}{\end{eqnarray}}

\allowdisplaybreaks
\DeclareGraphicsExtensions{.pdf,.pdf,.png,.jpg,.jpeg}

\newcommand*{\LEinline}[1]%
{\todo[inline, size=\footnotesize]{#1}}

\begin{document}

\preprint{MSUHEP-19-020, PITT-PACC-1905, SMU-HEP-19-14}

\title{Progress in the CTEQ-TEA NNLO global QCD analysis}

\author{Tie-Jiun Hou\footnote{E-mail: houtiejiun@mail.neu.edu.cn}${}^1$, Keping Xie${}^{2,9}$, Jun Gao${}^{3,4}$, Sayipjamal Dulat${}^5$, Marco Guzzi${}^6$, \linebreak T.~J. Hobbs${}^{2,7}$, Joey Huston${}^8$, Pavel Nadolsky\footnote{E-mail:nadolsky@smu.edu}\ ${}^2$, Jon Pumplin${}^8$, Carl Schmidt${}^8$, Ibrahim Sitiwaldi${}^5$, Dan Stump${}^8$, C.-P. Yuan\footnote{E-mail: yuan@pa.msu.edu}${}^8$}

\affiliation{
	${}^1$Department of Physics, College of Sciences, Northeastern University, Shenyang 110819, China\\
${}^2$Southern Methodist University, Dallas, TX 75275-0175, U.S.A.\\
	${}^3$Shanghai Jiao Tong University, Shanghai 200240, China \\
	${}^4$Center for High Energy Physics, Peking University, Beijing 100871, China \\
	${}^5$Xinjiang University, Urumqi, Xinjiang 830046, China\\
	${}^6$Kennesaw State University, GA 30144 , U.S.A.\\
${}^{7}$Jefferson Lab, EIC Center, Newport News, VA 23606, U.S.A.\\
	${}^8$Michigan State University, East Lansing, MI 48824, U.S.A.\\
	${}^9$Department of Physics and Astronomy, University of Pittsburgh, Pittsburgh, PA 15260, USA
}

\date{August 29, 2019}

\begin{abstract}
We present the new CTEQ-TEA global analysis of quantum chromodynamics (QCD). In this analysis, parton distribution functions (PDFs) of the nucleon are determined within the Hessian method at the next-to-next-to-leading order (NNLO) in perturbative QCD, based on the most recent measurements from the Large Hadron Collider (LHC) and a variety of world collider data. Because of difficulties in fitting both the ATLAS 7 and 8 TeV $W$ and $Z$ vector boson production cross section data, and to examine the range of PDFs obtained with different factorization scales in deeply-inelastic scattering, we present four families of (N)NLO CTEQ-TEA PDFs, named CT18, A, X and Z PDFs, respectively. The provided CT18 PDFs are suitable for a wide range of applications at the LHC and in other experiments. 
\end{abstract}

\maketitle
\section{The CT18 family NNLO parton distribution functions \label{sec:CP}}
In this contribution, we provide an update on the development of a new generation of parton distribution functions (PDFs) from the CTEQ-TEA group. The global QCD analysis reported here was completed in 2018, and the resulting PDF families are designated as ``CT18''. A comprehensive discussion of the panoply of results obtained in the CT18 global analysis will be provided in an upcoming publication. In this short report, we summarize the key results and comparisons. The CT18 PDFs can be downloaded for public use from the CTEQ-TEA PDF website~\cite{ct18website}.

The CT18 parton distribution functions (PDFs) replace those of CT14 presented in Ref.~\cite{Dulat:2015mca} and the transitional CT14HERA2 fit \cite{Hou:2016nqm}.They include a variety of new LHC data, involving inclusive jet production, $W$, $Z$ and Drell-Yan production, and the production of top quark pairs, from ATLAS, CMS and LHCb, while retaining crucial {\it legacy} data, such as measurements from the Tevatron and the HERA Run I and Run II combined data. Measurements of processes in similar kinematic regions, by ATLAS and by CMS, allow crucial cross-checks of the data. Measurements by LHCb often allow extrapolations into new kinematic regions not covered by the other experiments. In addition to the default CT18 PDF set suitable for the majority of computations, we provide three complementary PDF sets CT18Z, A, and X that are obtained under modified assumptions and result in a somewhat different PDF behavior, as discussed in the second part of this report. 

The goal of the CT18 analysis is to include as wide a kinematic range for each measurement as allowed by reasonable agreement between data and theory. For the ATLAS 7 TeV jet data~\cite{Aad:2014vwa}, for example, all rapidity intervals can not be simultaneously used without the use of systematic error decorrelations provided by the ATLAS experiment. 
Even with the ATLAS-recommended decorrelations, the resultant $\chi^2$ is not optimal, resulting in less effective PDF constraints.
Inclusive cross section measurements for jet production have been carried out for two different jet radii by both ATLAS and CMS. For both experiments, we have chosen the data with the larger $R$-value, as the next-to-next-to-leading order (NNLO) prediction should have a higher accuracy. We evaluate the jet cross section predictions using a scale of $p_T^{jet}$, consistent with past usage at the next-to-leading order (NLO). The result is largely consistent with similar evaluations using a scale of $H_T$~\cite{Ridder:2015dxa}.

\begin{figure}[tb]
	\center
	\includegraphics[width=0.43\textwidth]{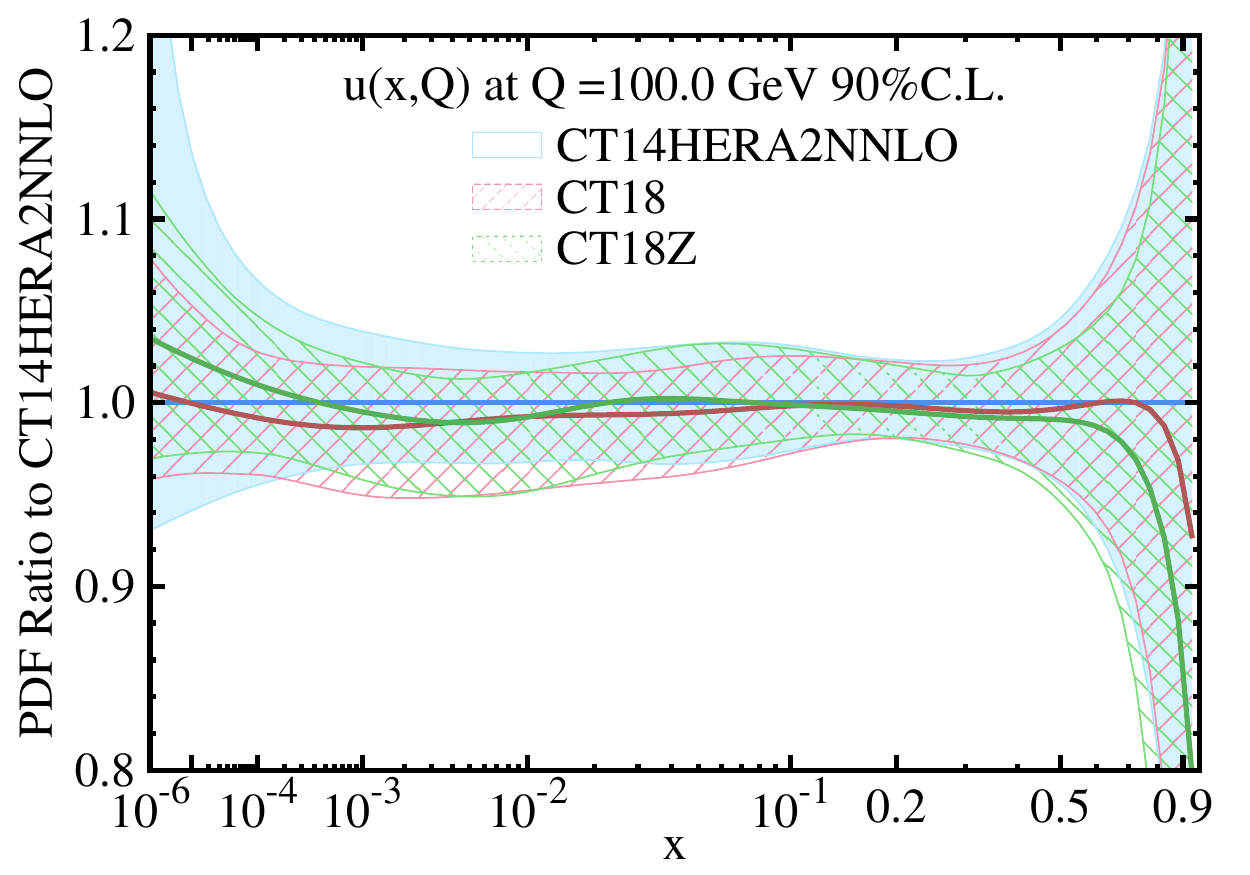}
	\includegraphics[width=0.43\textwidth]{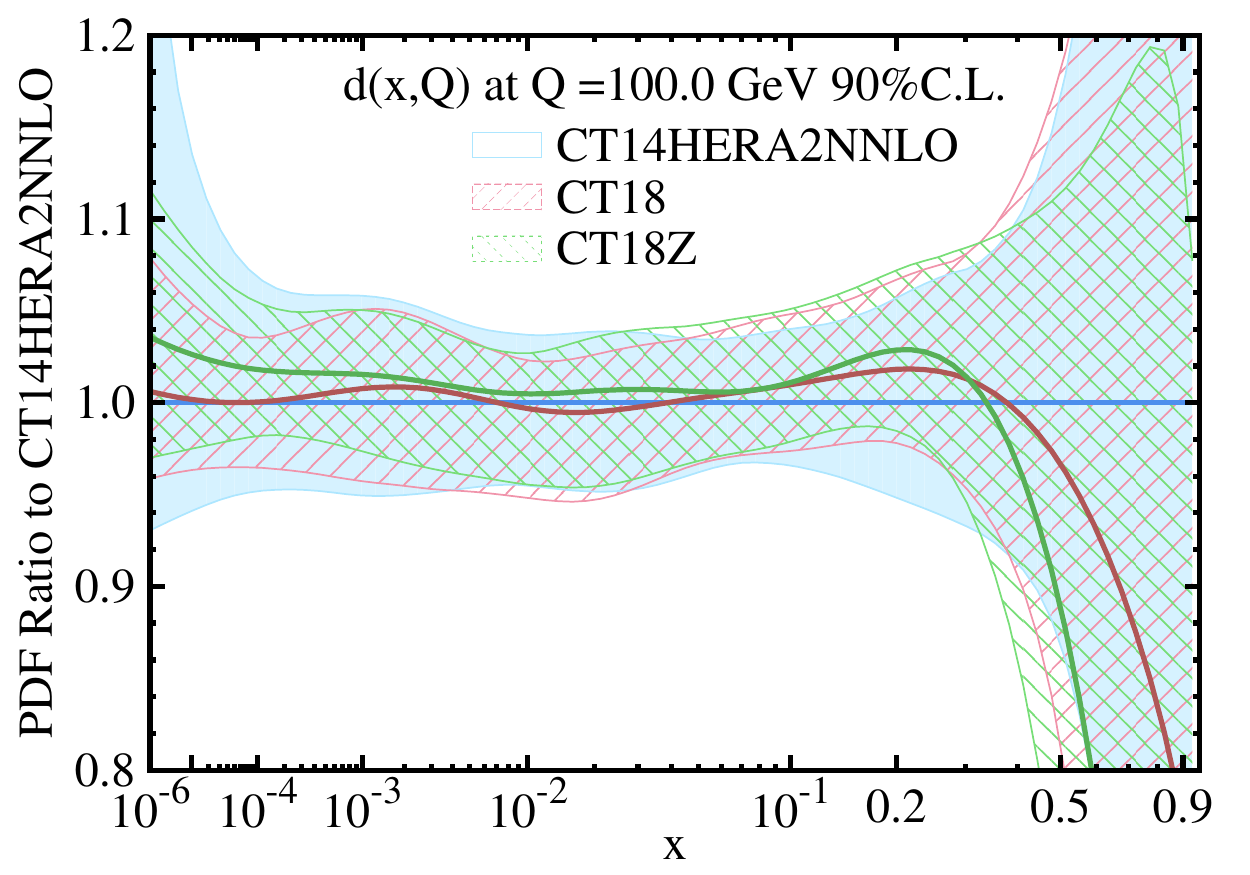}
	\includegraphics[width=0.43\textwidth]{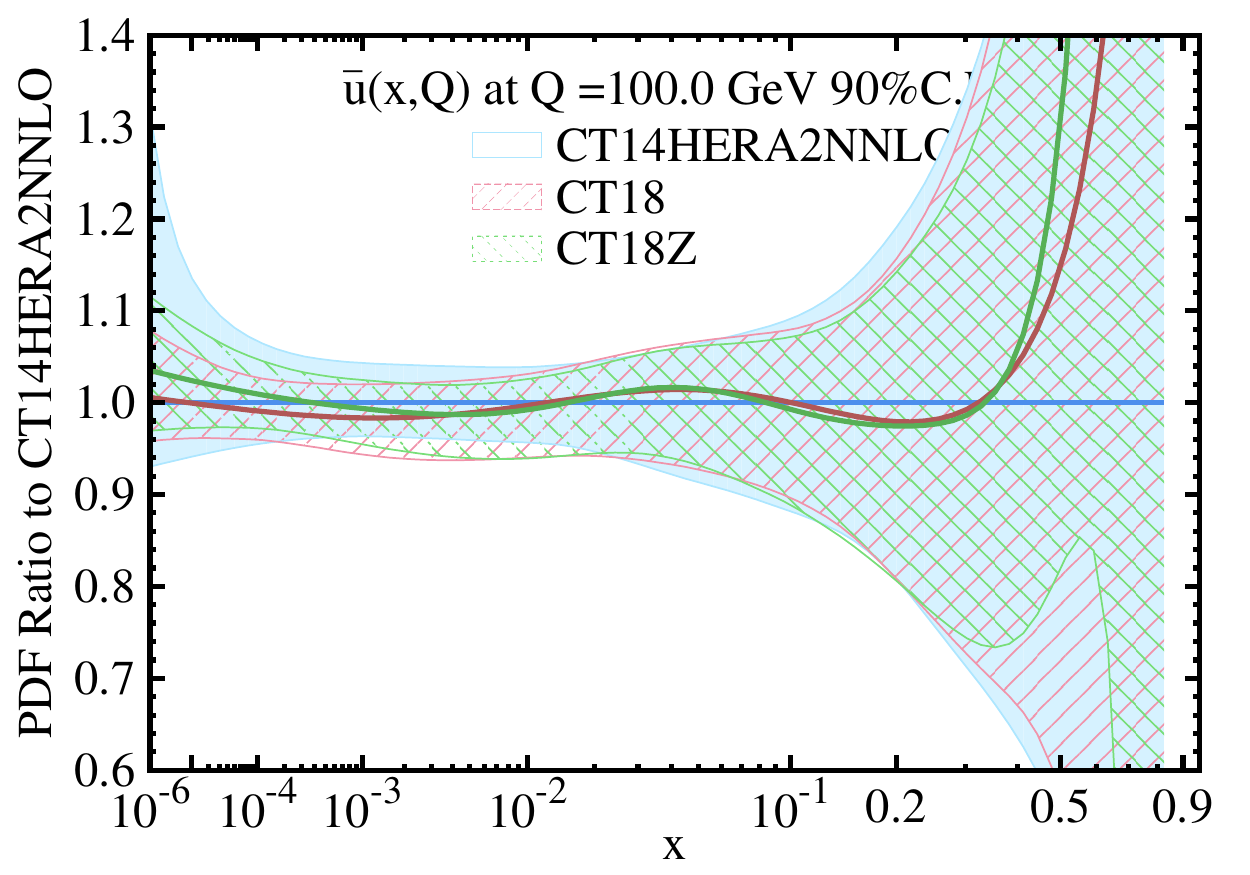}
	\includegraphics[width=0.43\textwidth]{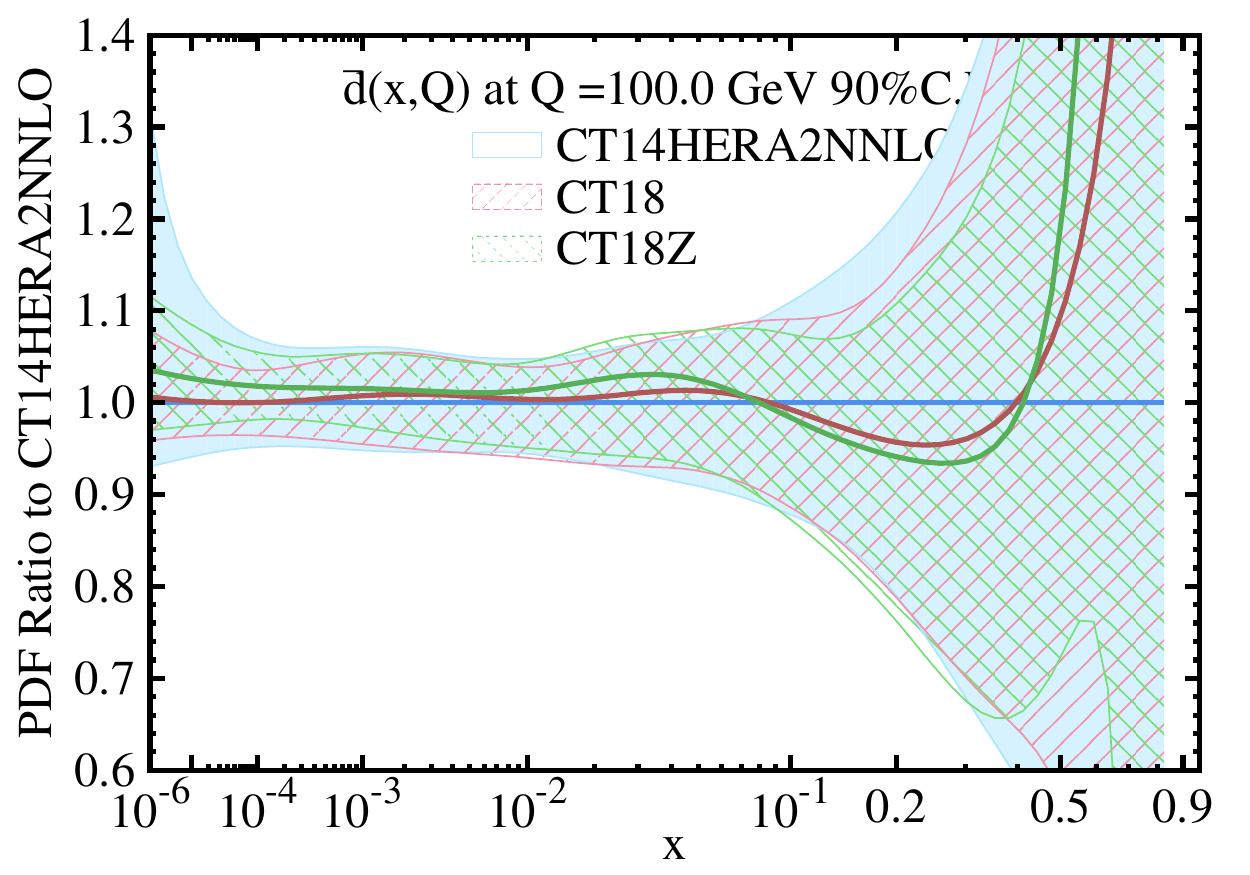}
	\includegraphics[width=0.43\textwidth]{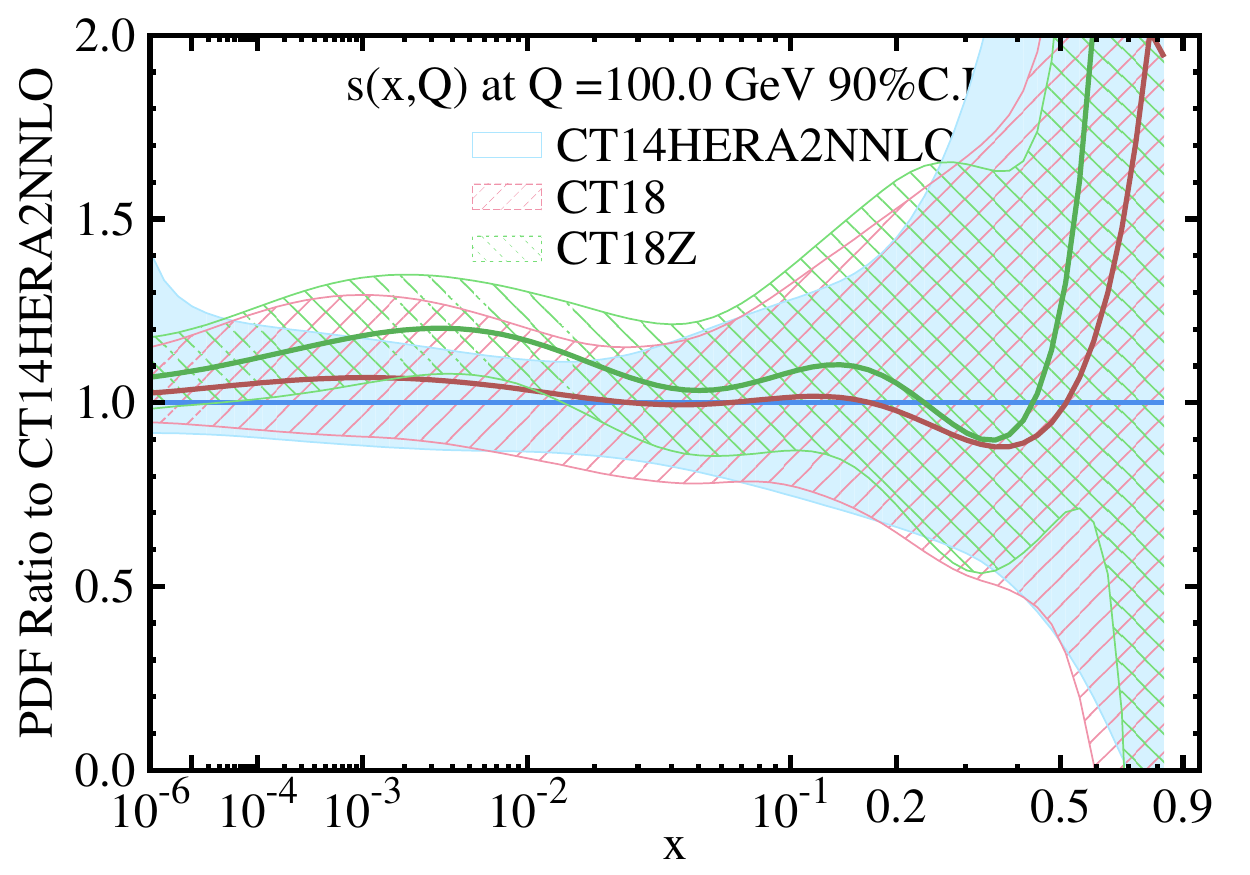}
	\includegraphics[width=0.43\textwidth]{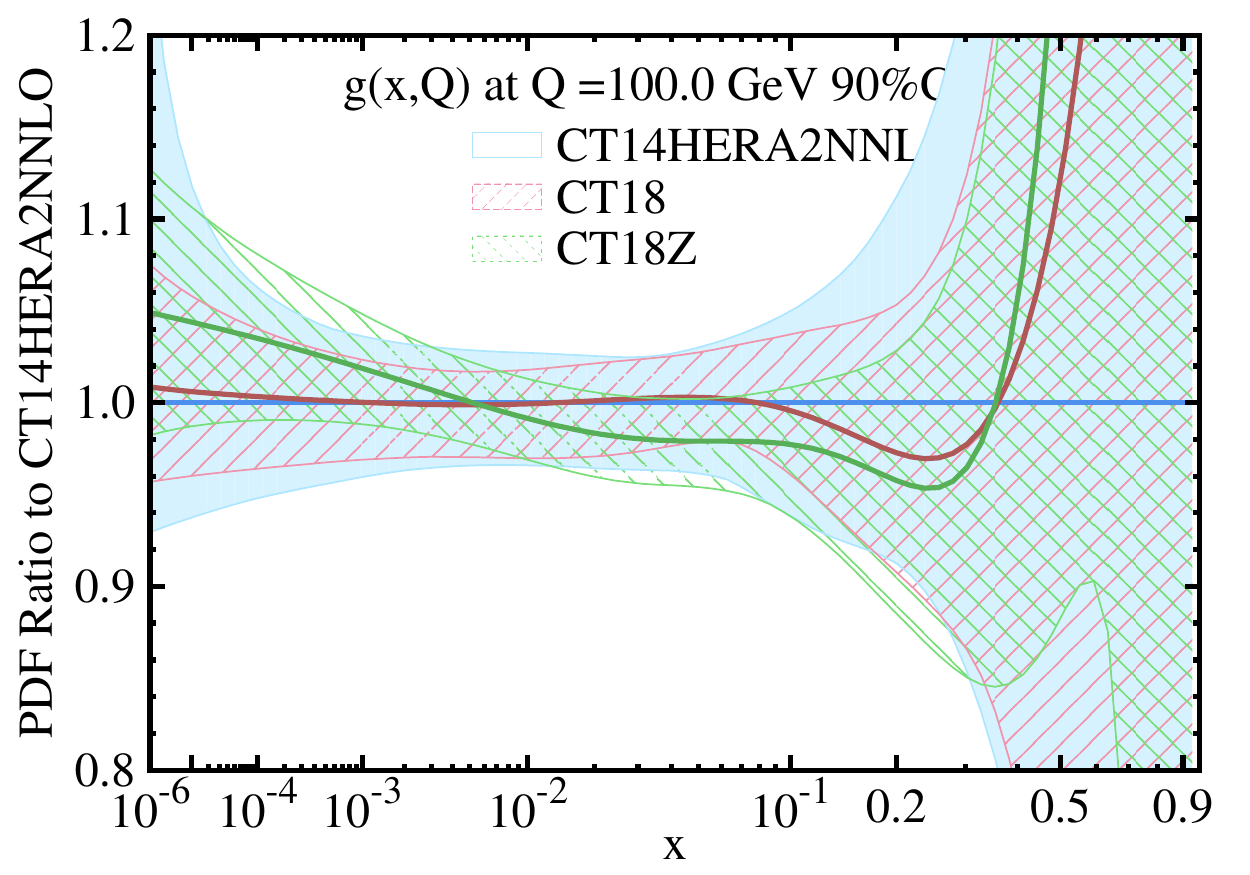}
	\caption{A comparison of 90\% C.L. PDF uncertainties from CT18  (red curve), CT18Z (green curve) and CT14HERA2 (blue curve) NNLO error ensembles at $Q=100$ GeV. The error bands are
		normalized to the respective central CT14HERA2 NNLO PDFs.
		\label{fig:PDFbands1}}
\end{figure}

In an ideal world, all such data sets would perfectly be compatible with each other, but differences are observed that do result in some tension between data sets and pulls in opposite directions. One of the crucial aspects of carrying out a global PDF analysis is dealing with data sets that add some tension to the fits, while preserving the ability of the combined data set to improve on the existing constraints on the PDFs. In some cases, a data set may be in such tension as to require its removal from the global analysis, or its inclusion only in a separate iteration of the new PDF set. Later, we will describe how the high precision ATLAS $W$ and $Z$ rapidity distributions \cite{Aaboud:2016btc} require the latter treatment.

Theoretical predictions for comparison to the data used in the global fit have been carried out at NNLO, either indirectly through the use of fast interpolation tables such as NLO \textsc{ApplGrid}~\cite{Carli:2010rw} and NNLO/NLO $K$-factors, or directly (for top-quark related observables) through the use of \textsc{fastNNLO} grids~\cite{Czakon:2017dip,Wobisch:2011ij}. New flexible PDF parametrizations have been tested for CT18, to minimize any parametrization bias. In some kinematic regions, there are few constraints from the data on certain PDFs. Lagrange Multiplier constraints are then applied to limit those PDFs to physically reasonable values. 
A 0.5\% uncorrelated error is included to account for numerical uncertainties in the Monte Carlo integration of NNLO cross sections of (i) ATLAS 7 TeV ~\cite{Aad:2014vwa} and CMS 7~\cite{Chatrchyan:2014gia} and 8 TeV~\cite{Khachatryan:2016mlc} jet productions; and (ii) ATLAS 8 TeV high-$p_T$ $Z$ production~\cite{Aad:2015auj}.

CT18 analysis includes new LHC experiments on $W$, $Z$, Drell-Yan, high-$p_T$ $Z$, jet, and $t \bar t$ pair productions, up to 30 candidate LHC data sets. 
The alternative CT18Z fit contains the following variations from the CT18 fit: (i) add in the ATLAS 7 TeV 4.6 fb$^{-1}$, $W$ and $Z$ rapidity distribution measurement~\cite{Aaboud:2016btc} which is not included in the CT18 fit, (ii) remove the CDHSW data, (iii) take charm pole mass to be 1.4 GeV, instead of the nominal value of 1.3 GeV, (iv) use a saturation scale, instead of the nominal scale of $Q$, for all the deep-inelastic scattering (DIS) processes in the fit.  
The final CT18(Z) data ensemble contains a total of 3681(3493) number of data points and $\chi^2/N_{pt}=1.17 (1.19)$ at the NNLO. 

The relative changes between the CT14HERA2 
NNLO \cite{Hou:2016nqm} and CT18 NNLO ensembles are best visualized by comparing
their PDF uncertainties. Fig.~\ref{fig:PDFbands1}
compares the PDF error bands at 90\% confidence level (CL)
for the key flavors, with each band normalized to the respective
best-fit CT14HERA2 NNLO PDF. The blue and red error bands are
obtained for CT14HERA2 NNLO PDFs and CT18 at $Q=100$ GeV, respectively.

\begin{figure}[tbp]
	\center
	\includegraphics[width=0.43\textwidth]{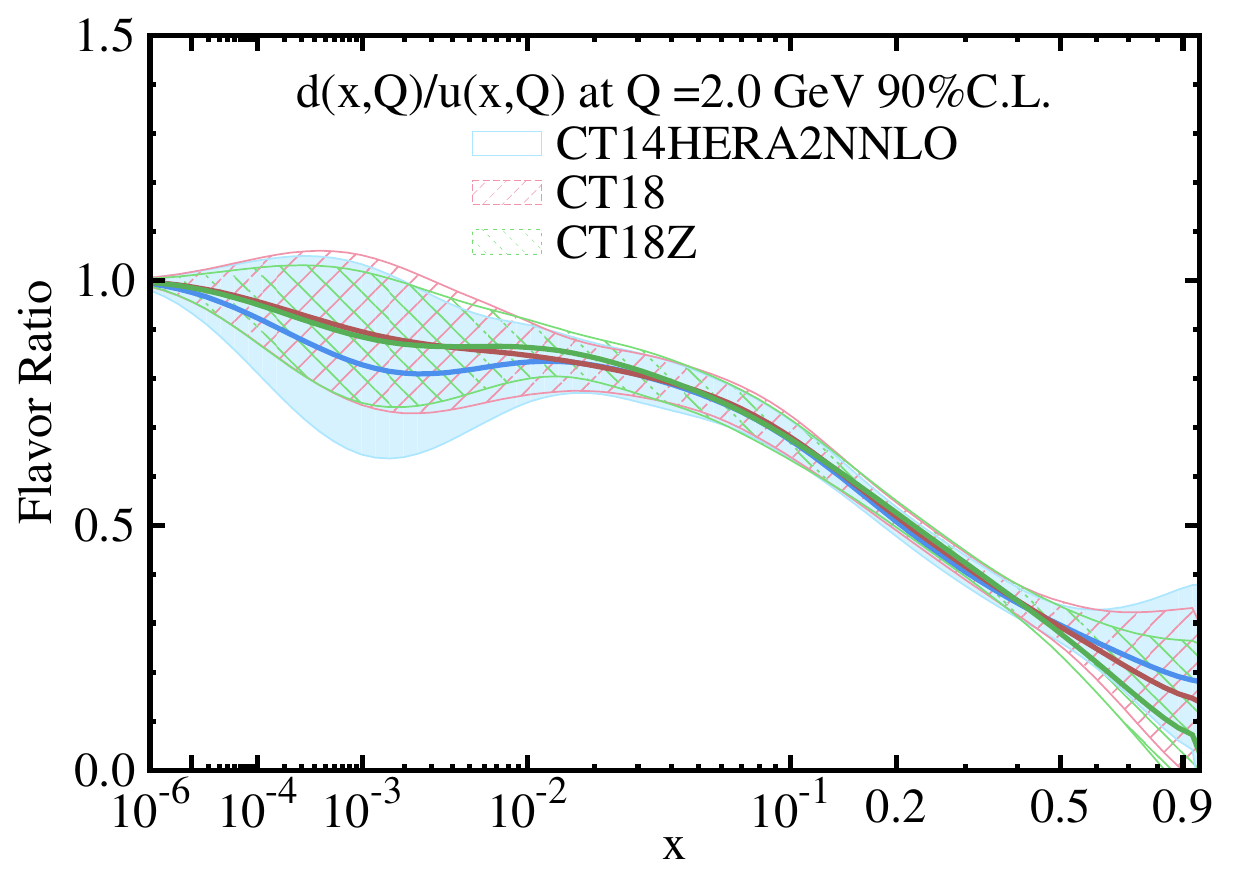}
	\includegraphics[width=0.43\textwidth]{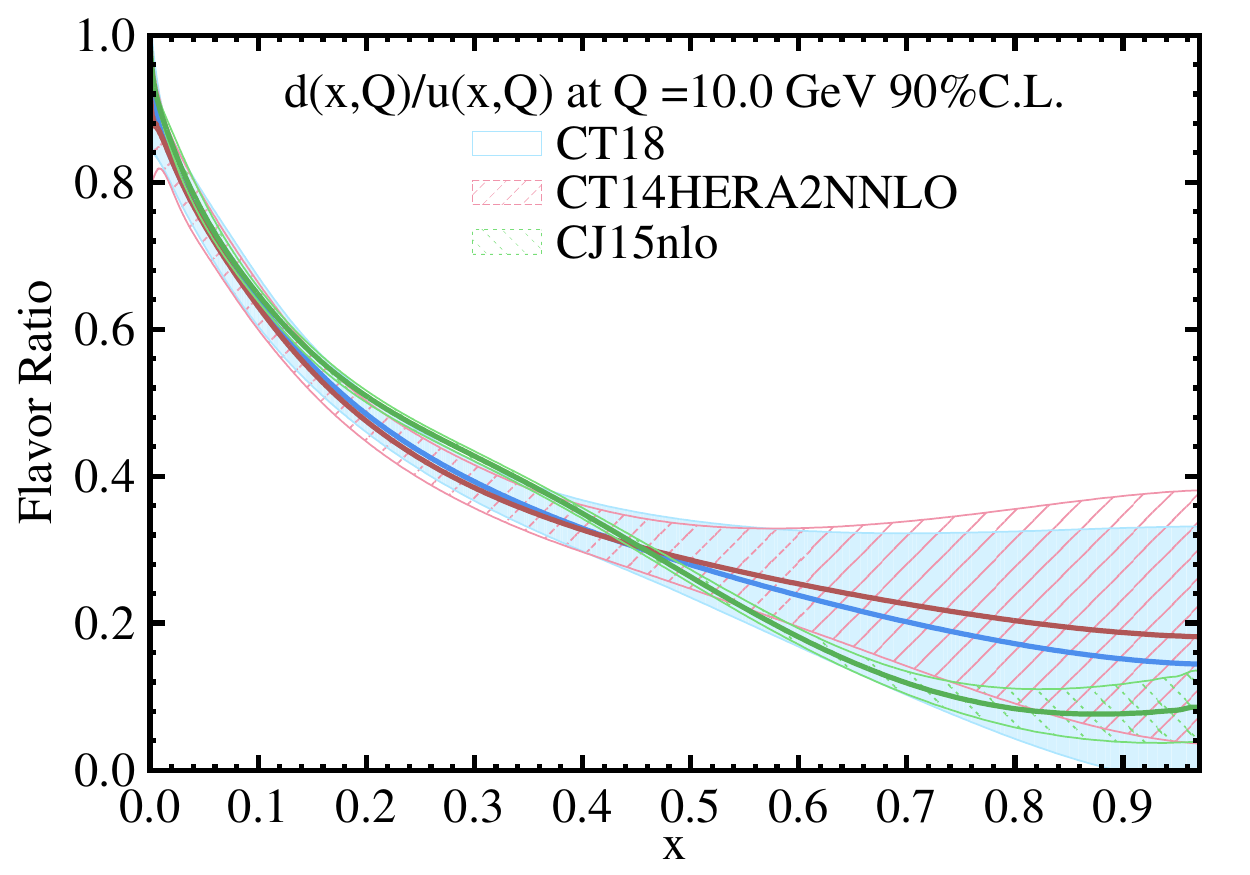}
	\caption{A comparison of 90\% C.L. uncertainties on the ratio
		$d(x,Q)/u(x,Q)$ for CT18  (red curve), CT18Z (green curve) and CT14HERA2 (blue curve) NNLO error ensembles at $Q=2$ or $10$ GeV, respectively.
		\label{fig:DOUband}}
\end{figure}

\begin{figure}[tb]
	\center
	\includegraphics[width=0.43\textwidth]{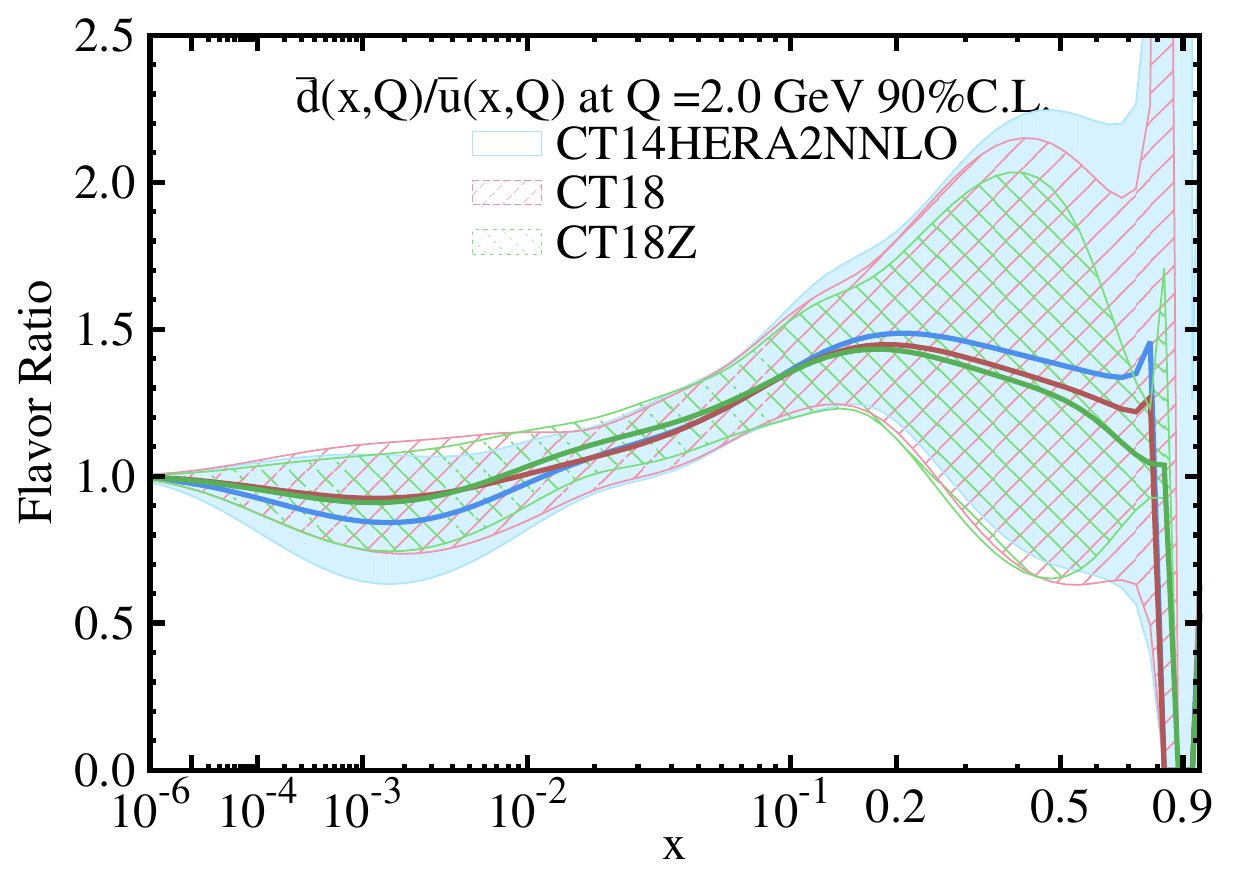}
	\includegraphics[width=0.43\textwidth]{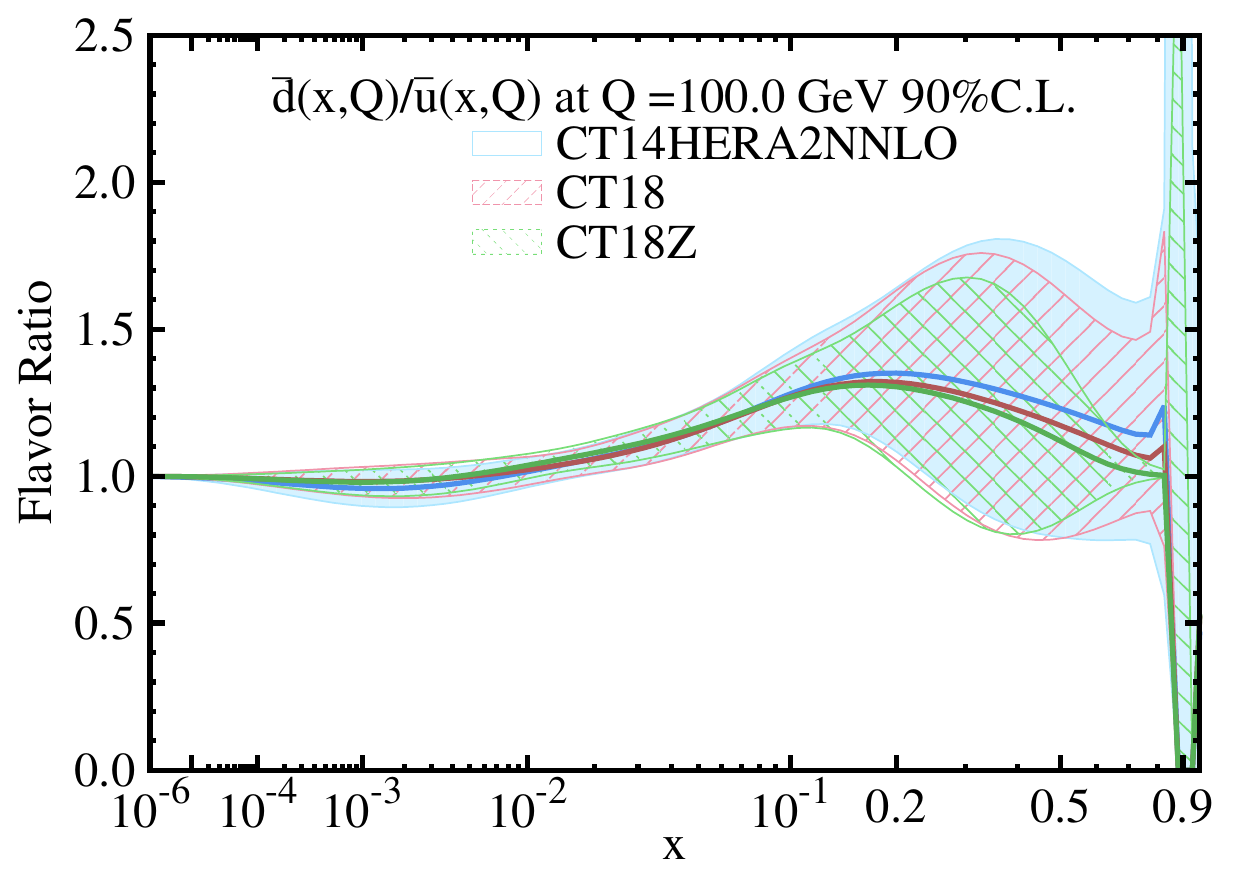}
	\includegraphics[width=0.43\textwidth]{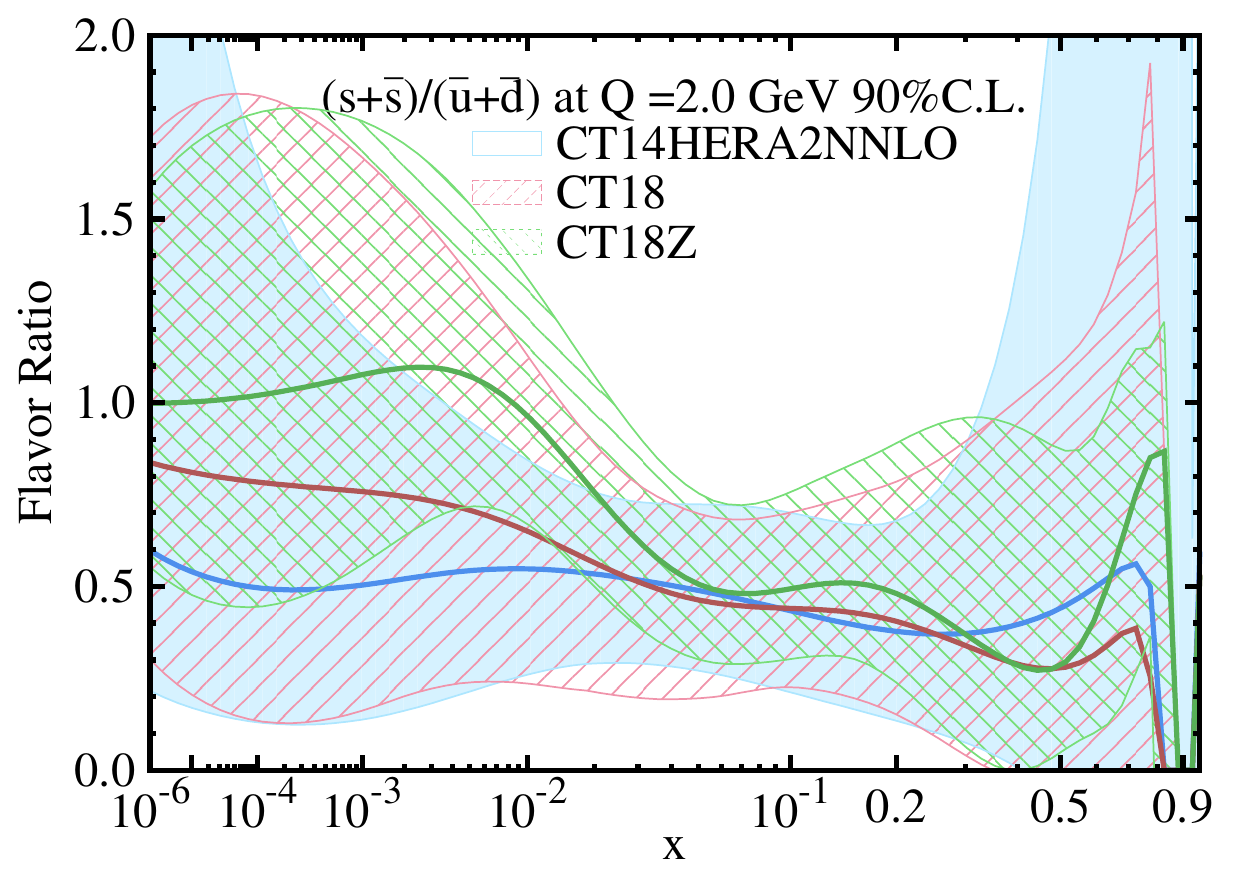}
	\includegraphics[width=0.43\textwidth]{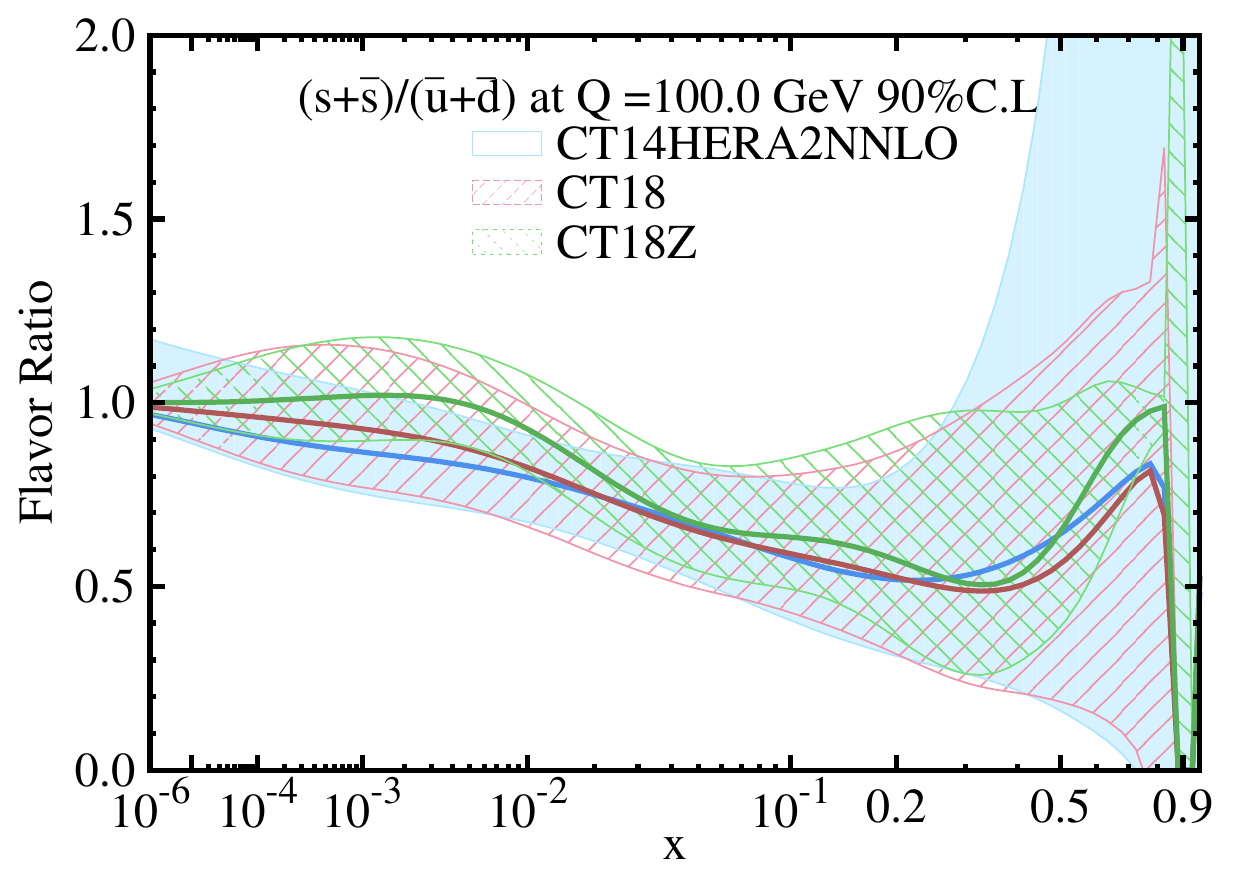}
	\caption{A comparison of 90\% C.L. uncertainties on the ratios
		$\bar d(x,Q)/\bar u(x,Q)$ and $\left(s(x,Q)+\bar
		s(x,Q)\right)/\left(\bar u(x,Q) +\bar d(x,Q)\right)$, for CT18  (red curve), CT18Z (green curve) and CT14HERA2 (blue curve) NNLO error ensembles at $Q=2$ or $100$ GeV, respectively.
		\label{fig:DBandSBbands}}
\end{figure}

Fig.~\ref{fig:DOUband} shows the different behaviors we find for the $d/u$ PDF ratio. The changes in $d/u$ in CT18, as compared
to CT14HERA2, can be summarized as a reduction of the central ratio at $x > 0.5$ and a
decreased uncertainty at $x < 10^{-3}$.
The collider charge asymmetry data constrains $d/u$ at
$x$ up to about 0.5. 
At even higher $x$, which is not directly constrained by the experiments we fit, the behavior of the CT18 PDFs reflects the parametrization form,
which now allows $d/u$ to
approach any constant value as $x\rightarrow 1$.
As noted earlier, the parametrization form of $u$, $d$, $\bar u$ and $\bar d$ quarks in CT18 are the same as those in CT14HERA2. 

\begin{figure}[t]
	\includegraphics[width=0.48\textwidth]{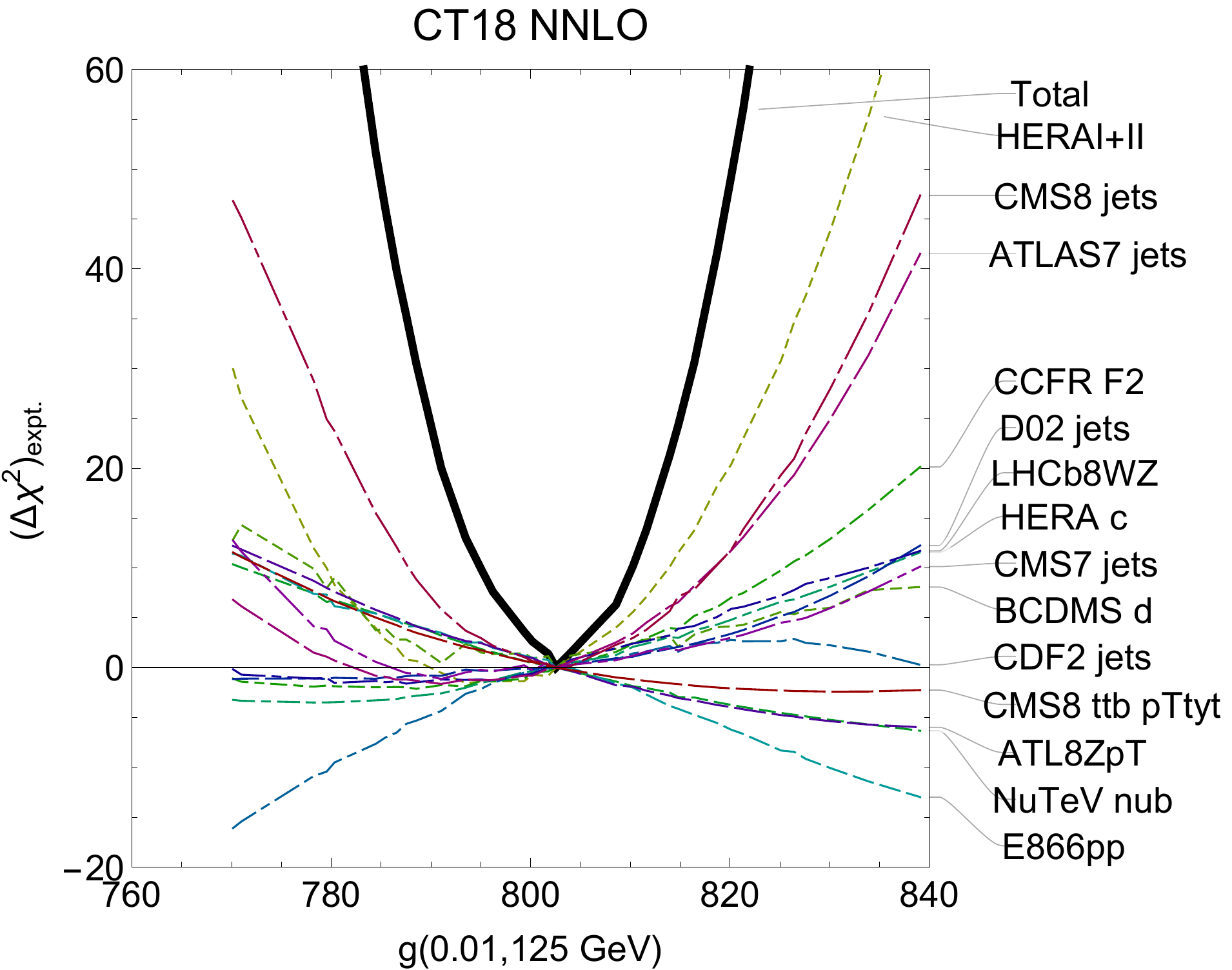}\quad
	\includegraphics[width=0.48\textwidth]{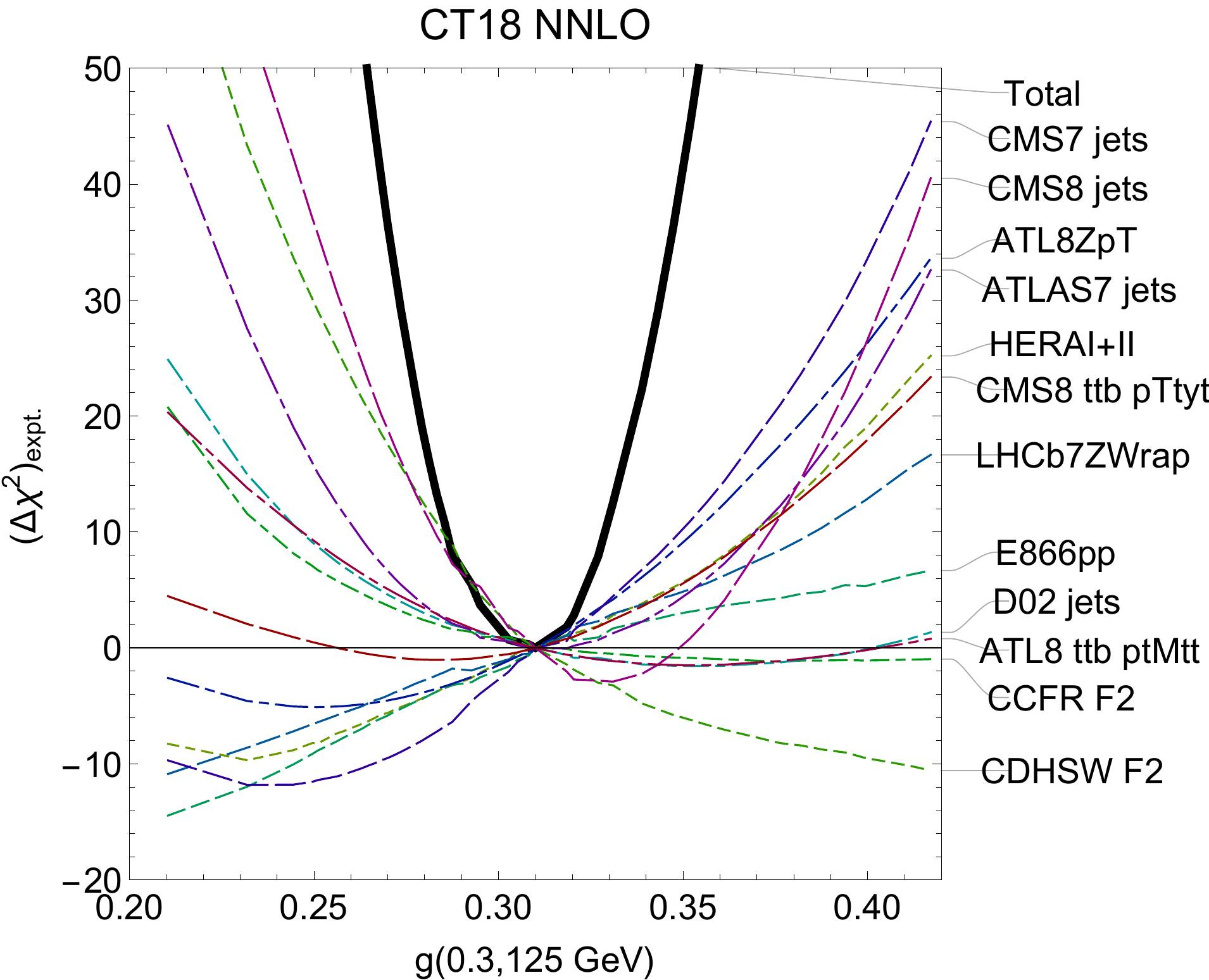}
	\caption{The Lagrange Multiplier scan of gluon PDF at $Q=125$ GeV and $x=0.01$ and $0.3$, respectively, for the CT18 NNLO fits.
		\label{fig:lm_g18}}
\end{figure}

\begin{figure}[t]
	\begin{center}
		\includegraphics[width=0.48\textwidth]{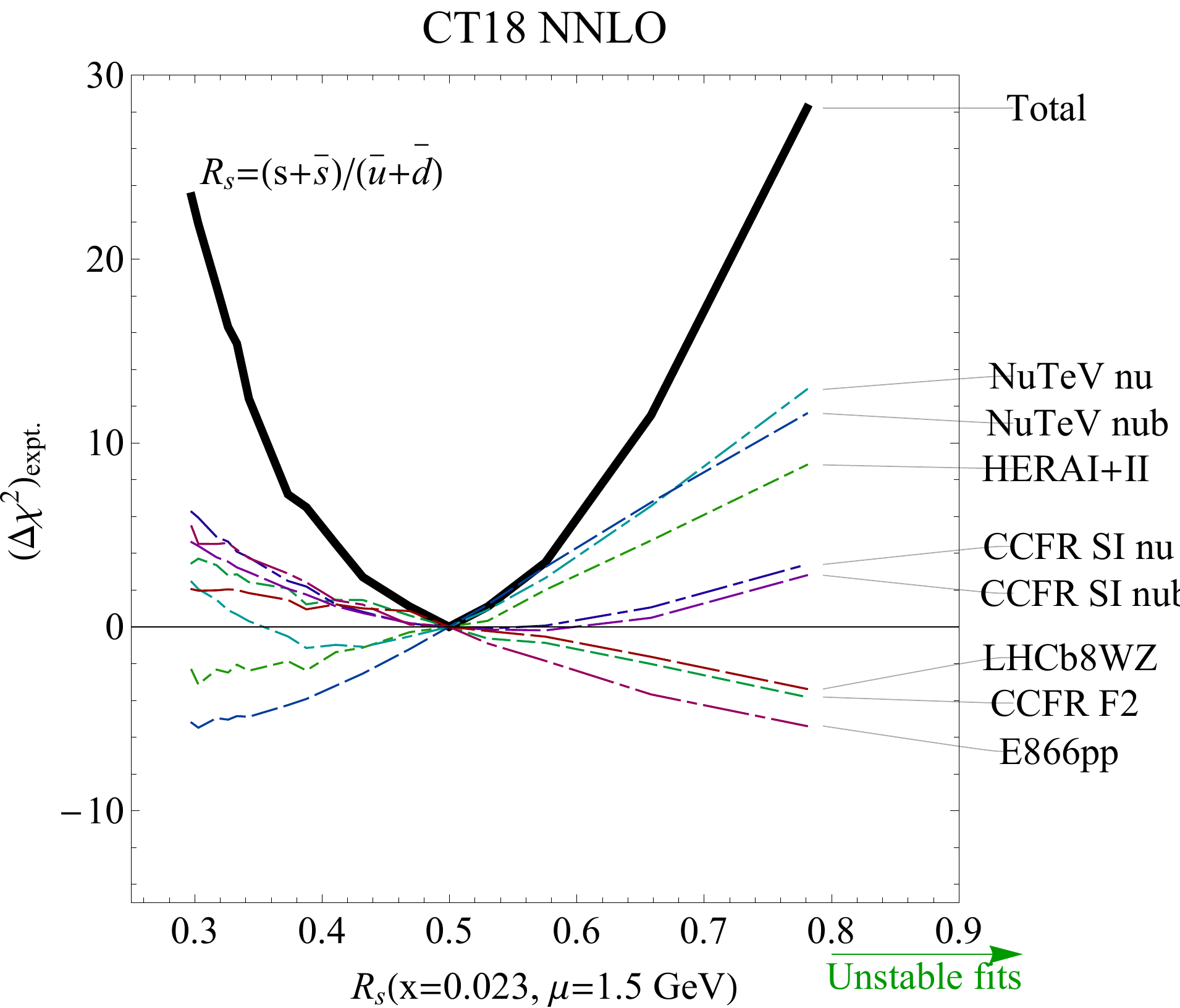}\quad
		\includegraphics[width=0.48\textwidth]{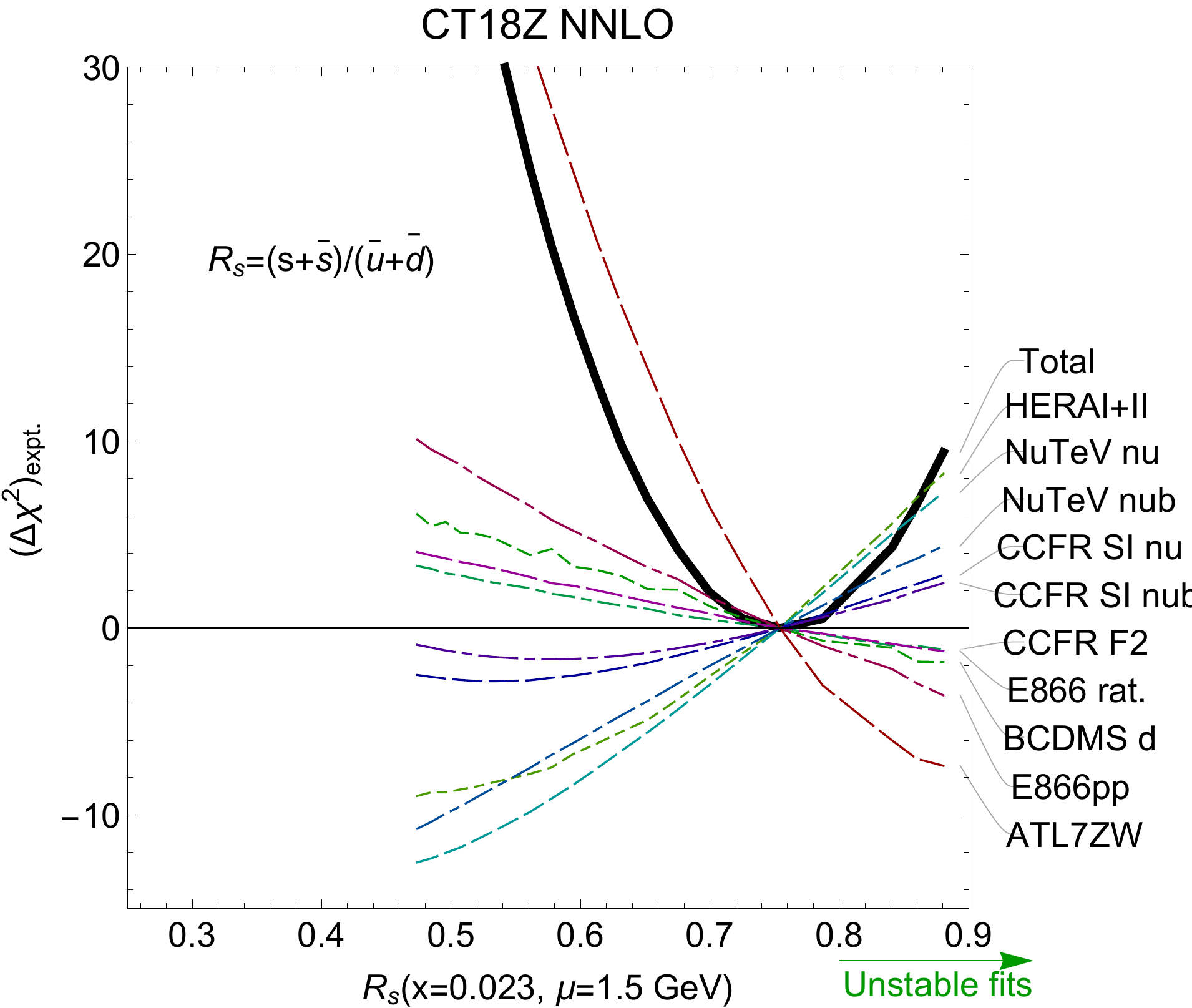}\quad
		\caption{The Lagrange Multiplier scan of $R_s$ at $Q=1.5$ GeV and $x=0.023$ for CT18, and CT18Z fits.
			\label{fig:lm_rs}}
	\end{center}
\end{figure}

Turning now to the ratios of sea quark PDFs in
Fig.~\ref{fig:DBandSBbands}, we observe that the uncertainty on $\bar
d(x,Q)/\bar u(x,Q)$ in the left inset has decreased at small $x$
in CT18. At $x > 0.1$, the CT18 non-perturbative parametrization forms for $\bar u$ and $\bar d$ ensure that the ratio
$\bar d(x,Q_0)/\bar u(x,Q_0)$, with $Q_0=1.3$ GeV, can approach a constant value that
comes out to be close to 1 in the central fit. The uncertainty on
$\bar d/\bar u$ has also decreased across most of the $x > 2 \times 10^{-3}$ range, especially around $x \sim 0.1$.

The overall increase in the strangeness PDF at $ x < 0.03$ and decrease of $\bar u$ and 
$\bar d$ PDFs at $ x  < 10 ^{-3}$ lead to a
larger ratio of the strange-to-nonstrange sea quark PDFs,
 $\left(s+\bar s\right)/\left(\bar u +\bar d \right)$, 
 presented in
Fig.~\ref{fig:DBandSBbands}. 
At $ x  < 10 ^{-3}$, this
ratio is determined entirely by parametrization form and was found
in CT10 to be consistent with the exact $SU(3)$ symmetry of PDF
flavors, 
$\left(s+\bar s\right)/\left(\bar u +\bar d\right) \rightarrow 1$
at $x\rightarrow 0$, albeit with a large uncertainty. 
The $SU(3)$-symmetric asymptotic solution at $x\rightarrow 0$
was not enforced in CT14, nor CT14HERA2, so that this ratio at $Q=2$ GeV 
is about 0.6 at $x=10^{-6}$. In CT18, we have taken a different $s$-PDF non-perturbative parametrization form and assumed the exact $SU(3)$ symmetry of PDF
flavors so that this ratio asymptotically approaches to 1 as $x \to 0$.

One technique for studying the parton PDFs is to compute Lagrange Multiplier scans with respect to some features of $f(x,Q)$. Two examples are shown here. 
In the first example, we study  constraints on the gluon-PDF at $Q=125$ GeV and $x=0.01$ and $0.3$, from various experimental data, cf. Fig.~\ref{fig:lm_g18}. 
In the second example, we show constraints on 
the $R_s \equiv (s+ \bar s)/({\bar u}+ {\bar d})$ ratio at $Q=1.5$ GeV, $x=0.023$ and $x=0.1$, cf. Fig.~\ref{fig:lm_rs}. 

\begin{figure}[tb]
	\begin{center}
		\includegraphics[width=0.45\textwidth]{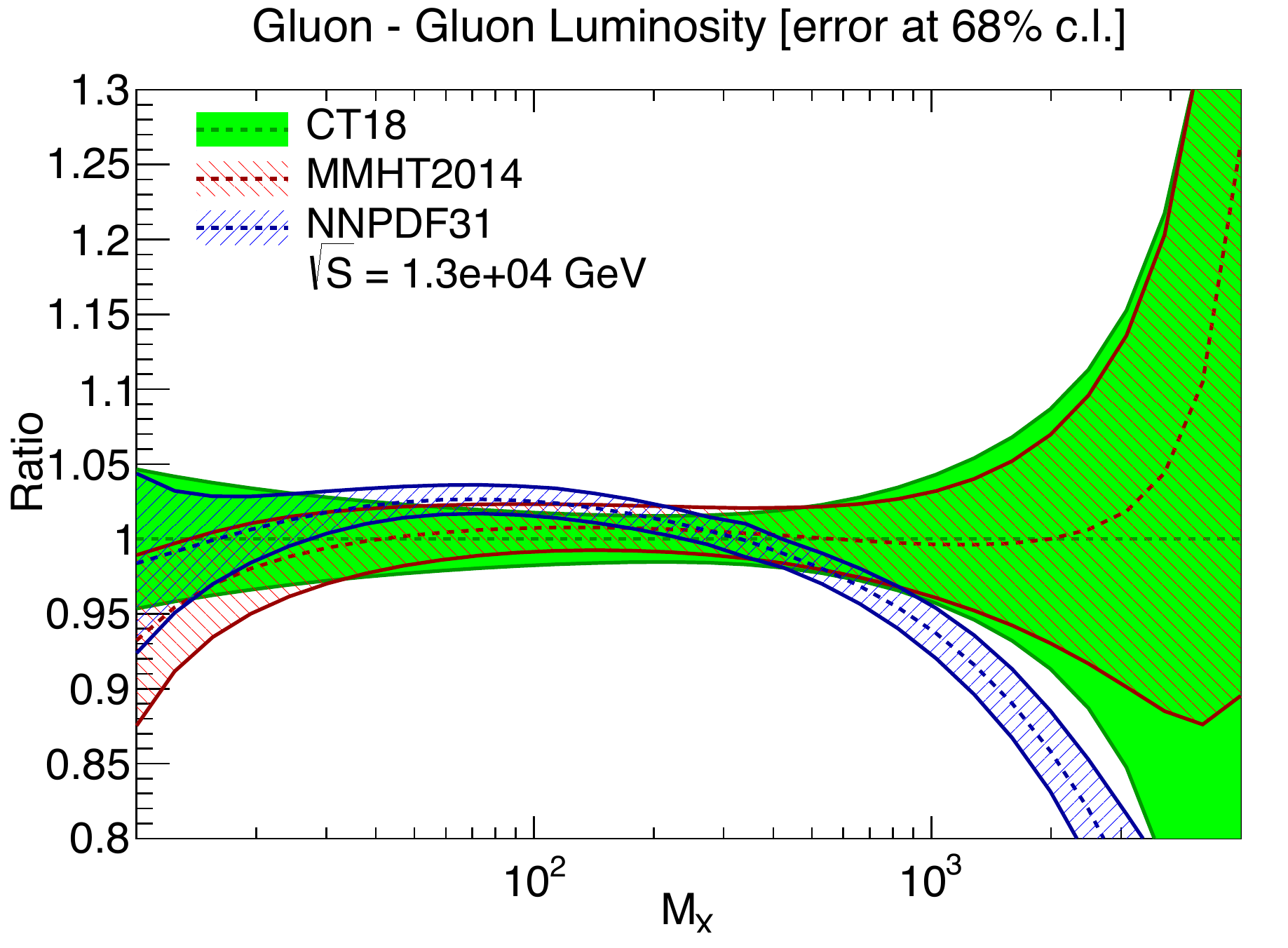}		\includegraphics[width=0.45\textwidth]{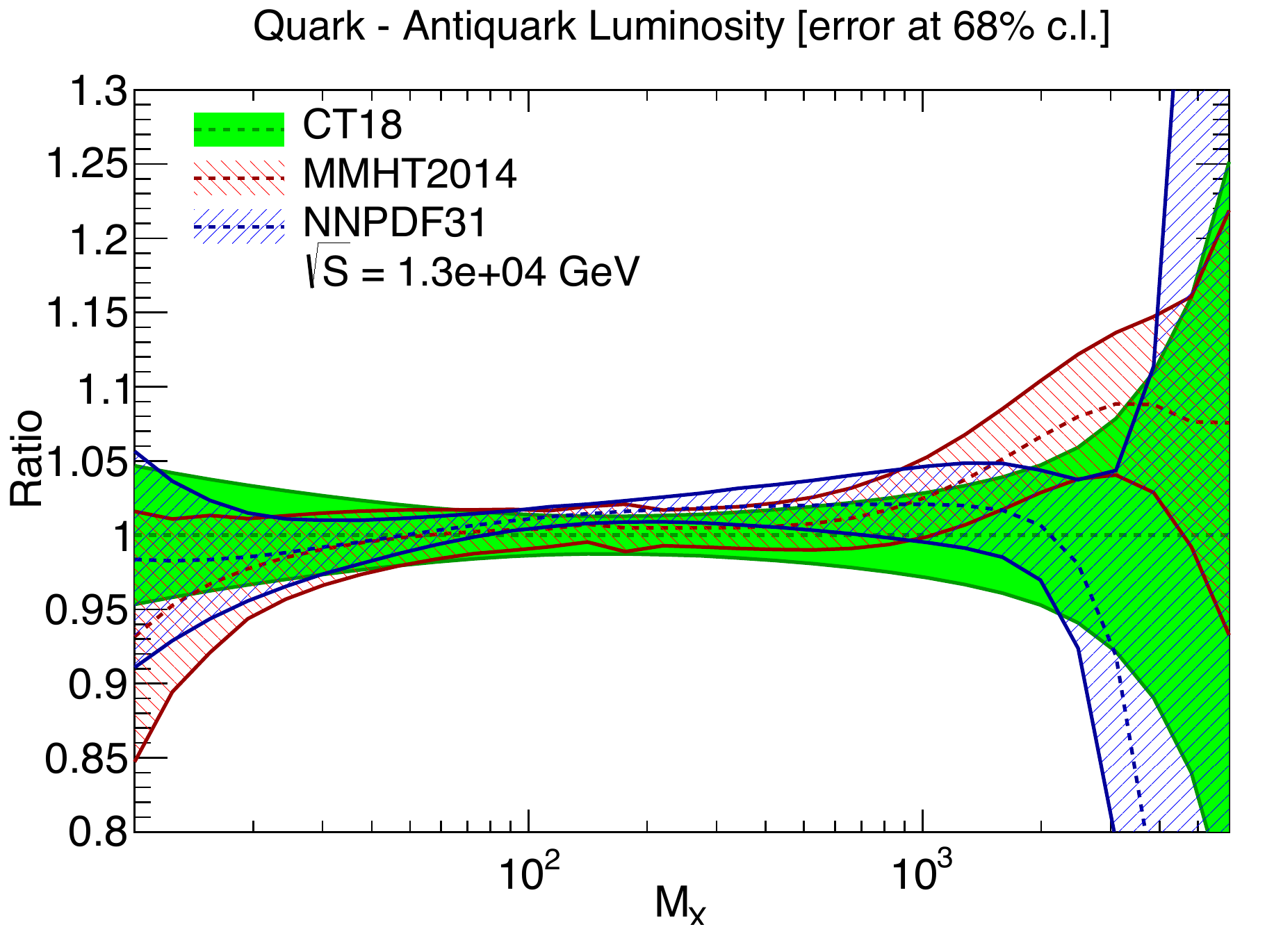}		
		\includegraphics[width=0.45\textwidth]{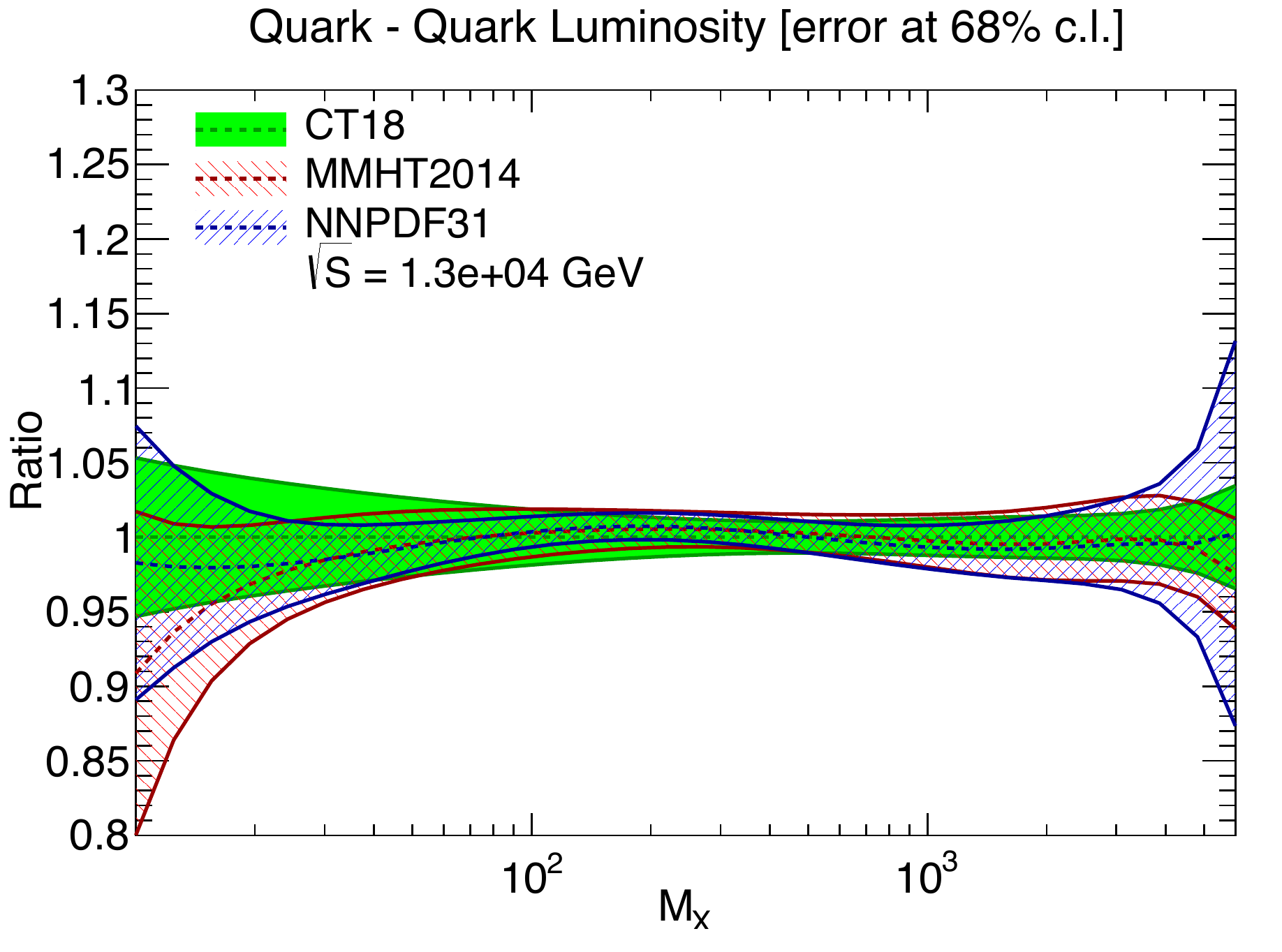}
		\includegraphics[width=0.45\textwidth]{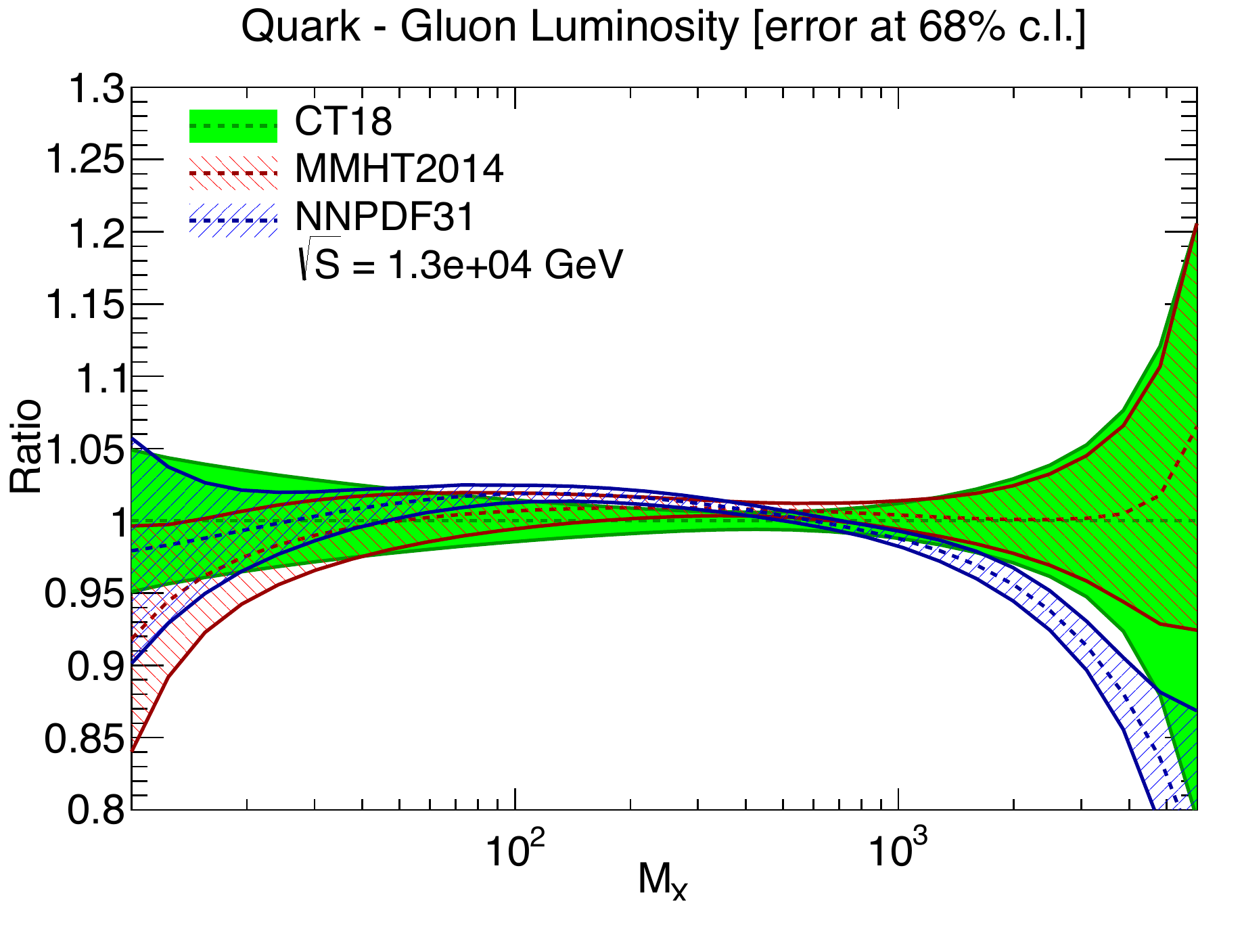}		
	\end{center}
	\vspace{-2ex}
	\caption{\label{fig:lumib}
		Comparison of various PDF luminosities at the 13 TeV LHC. 
	}
\end{figure}

Finally, we compare various PDF luminosities at the 13 TeV LHC, as shown in Fig.~\ref{fig:lumib}.

\clearpage
\section{LHC and DIS experimental data in the CT18(Z) global QCD analysis}

In the previous section, we presented the
new CT18 global QCD analysis of parton distribution functions
(PDFs). Let us now discuss in more detail the selection and implementation of the new data sets in the CT18 global fit, and the associated physics issues that 
affect the resulting PDFs and a wide class of QCD predictions
based on them. 

By 2018, the LHC collaborations published
about three dozen experimental data sets that can potentially constrain 
the CT PDFs. In light of the unprecedented precision reached in some 
measurements, the latest LHC data must be analyzed using
next-to-next-to-leading order (NNLO) theoretical predictions in
perturbative QCD. The final PDFs depend 
on numerous systematic factors in the experimental data; and the scope
of numerical computations needs to be expanded, too. A systematic examination 
of these effects is essential for trustworthy estimates of PDF uncertainties. 

\textbf{Combined HERA I+II DIS data and an $x$-dependent factorization
  scale.} Even in the 
LHC era, the DIS data from $ep$ collider HERA provides the dominant
constraints on the CT18 PDFs. This dominance can be established using
the \texttt{ePump} and \texttt{PDFSense} statistical techniques reviewed below.
CT18 implements the final (``combined'') data set from DIS at
HERA-I and II \cite{Abramowicz:2015mha} that supersedes the HERA-I
only data set \cite{Aaron:2009aa} used in CT14 \cite{Dulat:2015mca}.
A transitional PDF set, CT14HERA2, was released based on fitting the
final HERA data \cite{Hou:2016nqm}. We found fair overall agreement of the HERA I+II
data with both CT14 and CT14HERA2 PDFs, and that both PDF ensembles
describe equally well the non-HERA data included in our global analysis.
At the same time, we observed some disagreement (``statistical tension'') 
between the $e^{+}p$ and $e^{-}p$ DIS cross sections of the HERA I+II data set. 
We determined that, at the moment, no plausible explanation conclusively explains
the full pattern of these tensions, as they are distributed across
the whole accessible range of Bjorken $x$ and lepton-proton momentum
transfer $Q$ at HERA.

It has been argued that resummation of logarithms $\ln^p(1/x)$ at
$x\ll 1$ improves agreement withe HERA Run I+II data by several tens
of units of $\chi^2$ \cite{Ball:2017otu,Abdolmaleki:2018jln}. In our
analysis, we observe that, by evaluating the DIS cross sections at
NNLO with an $x$-dependent factorization scale, such as a tuned scale
$\mu^2_{F,x} = 0.8^2\ \left(Q^2 + 0.3\mbox{\, GeV}^2/x^{0.3}\right)$,
instead of the conventional choice $\mu^2_F =Q^2$, we achieve a
comparable quality of improvement in the description of the HERA DIS
data set by the {\it fixed-order} NNLO theoretical prediction as the
inclusion of the low-$x$ resummation in
\cite{Ball:2017otu,Abdolmaleki:2018jln}. Namely, the
$\chi^2(\mbox{HERAI+II})$ reduces by $> 50$ units in the kinematical region
$Q > 2\mbox{ GeV}$, $x > 10^{-5}$ of the DIS data included in the CT18
global fit. The parametric form of the $x$-dependent scale
$\mu^2_{F,x}$ is inspired by qualitative saturation arguments (see, e.g.,
\cite{Caola:2009iy}), and the numerical coefficients in $\mu^2_{F,x}$
are chosen to  minimize $\chi^2$ for the HERA DIS data.

\begin{figure}[tb]
	\includegraphics[width=0.59\textwidth]{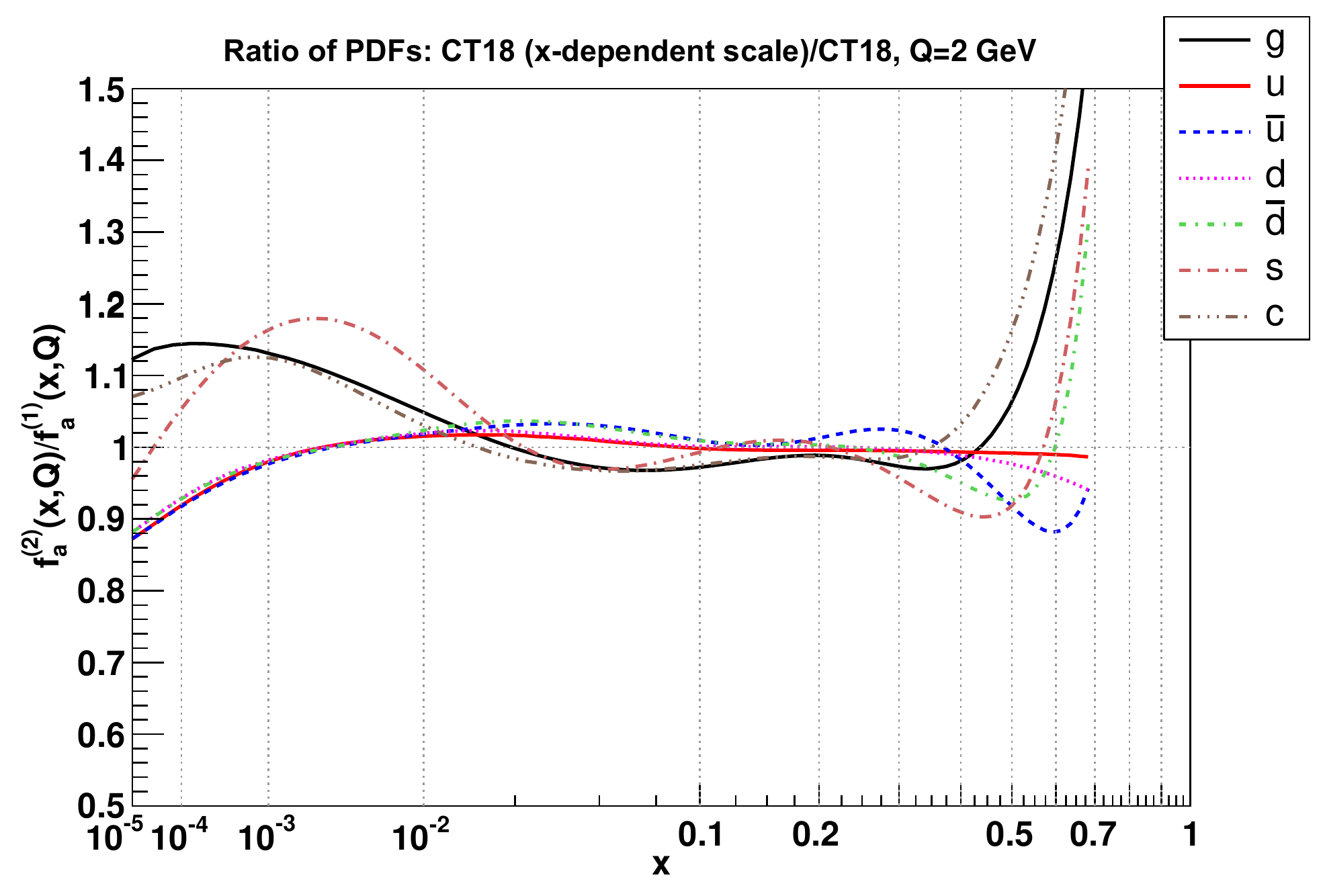}
	\includegraphics[width=0.4\textwidth]{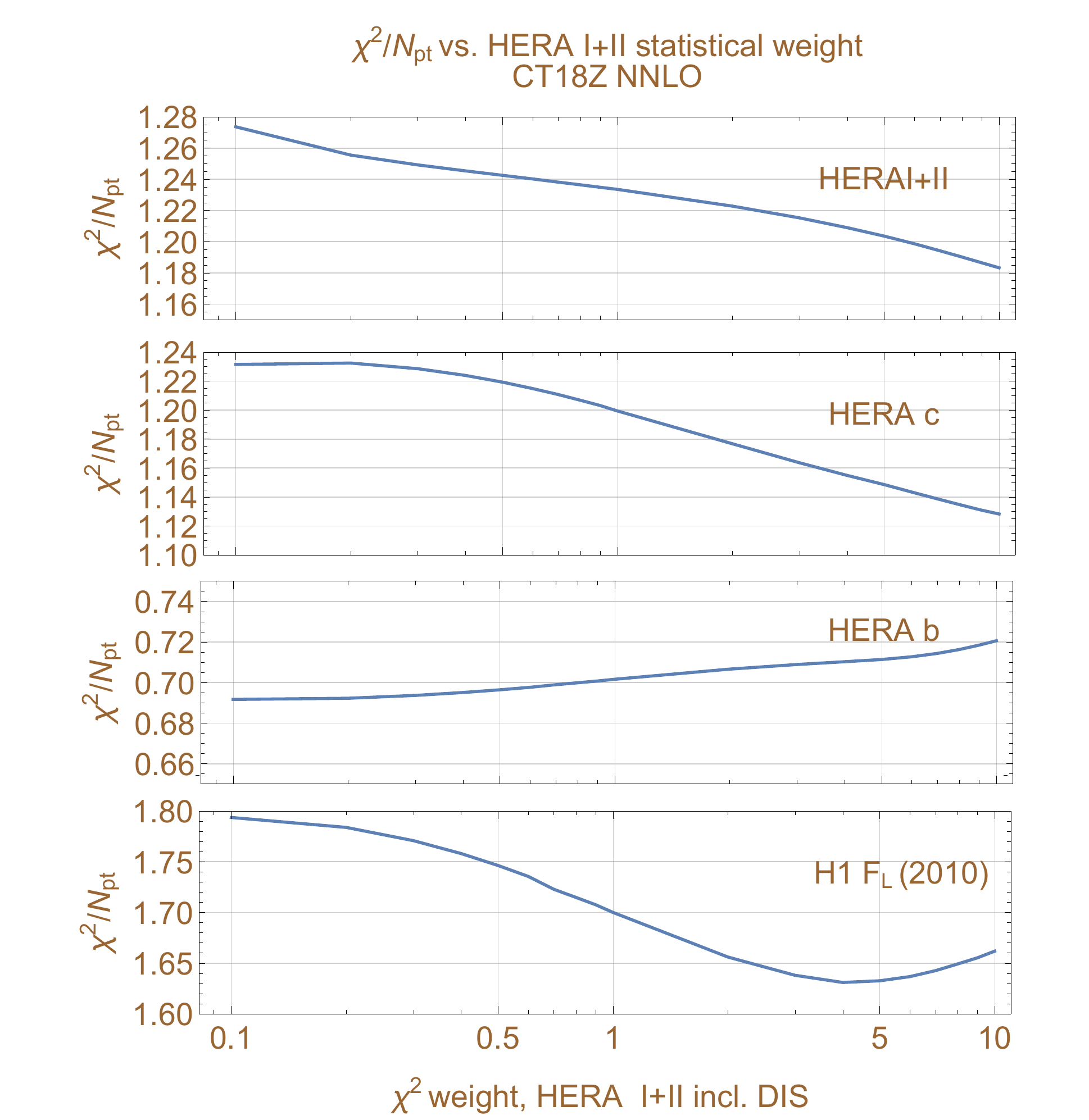}
\caption{Left: The ratios of the candidate CT18 NNLO PDFs obtained with the
  $x$-dependent and standard factorization scales in DIS data
  sets. Right: The $\chi^2/N_{pt}$ values for four HERA data sets in
  the CT18Z fits with the $x$-dependent DIS factorization scale and
  varied statistical weight of the HERA I +II inclusive DIS data set.}
\label{fig:saturation}
\end{figure}

Fig.~\ref{fig:saturation}(left) shows
the changes in the candidate CT18 PDFs obtained by fitting DIS with
the $x$-dependent factorization scale, as compared to the CT18 PDFs
with the nominal scale. With the scale $\mu_{F,x}^2$, we observe
reduced $u$ and $d$ (anti-)quark PDFs and increased gluon and
strangeness PDFs at $x < 10^{-2}$ as compared to the nominal CT18 fit,
with some compensating changes occuring in the same PDFs in the
unconstrained region $x > 0.5$ in order to satisfy the valence and
momentum sum rules. The right Fig.~\ref{fig:saturation} shows the
$\chi^2/N_{pt}$ values (divided by the number $N_{pt}$ of experimental data
points) for four HERA data sets (inclusive NC+CC DIS
\cite{Abramowicz:2015mha},
reduced charm, bottom production cross sections, and H1 longitudinal function
$F_L(x,Q^2)$ \cite{Collaboration:2010ry}) in the fits with the varied
statistical weight $w$ of the HERA I+II inclusive DIS data set \cite{Abramowicz:2015mha}. The
default CT18 fits correspond to $w=1$; with $w=10$, the CT18 fit
increasingly behaves as a HERA-only fit. We see that, with the scale
$\mu^2_{F,x}$ and $w=10$, $\chi^2/N_{pt}$ for the inclusive DIS data set
improves almost to the levels observed in the ``resummed'' HERA-only fits
without intrinsic charm \cite{Ball:2017otu,Abdolmaleki:2018jln}. The
quality of the fit to the charm SIDIS cross section and H1 $F_L$ also improves.

\begin{figure}[tb]
	\includegraphics[width=0.49\textwidth]{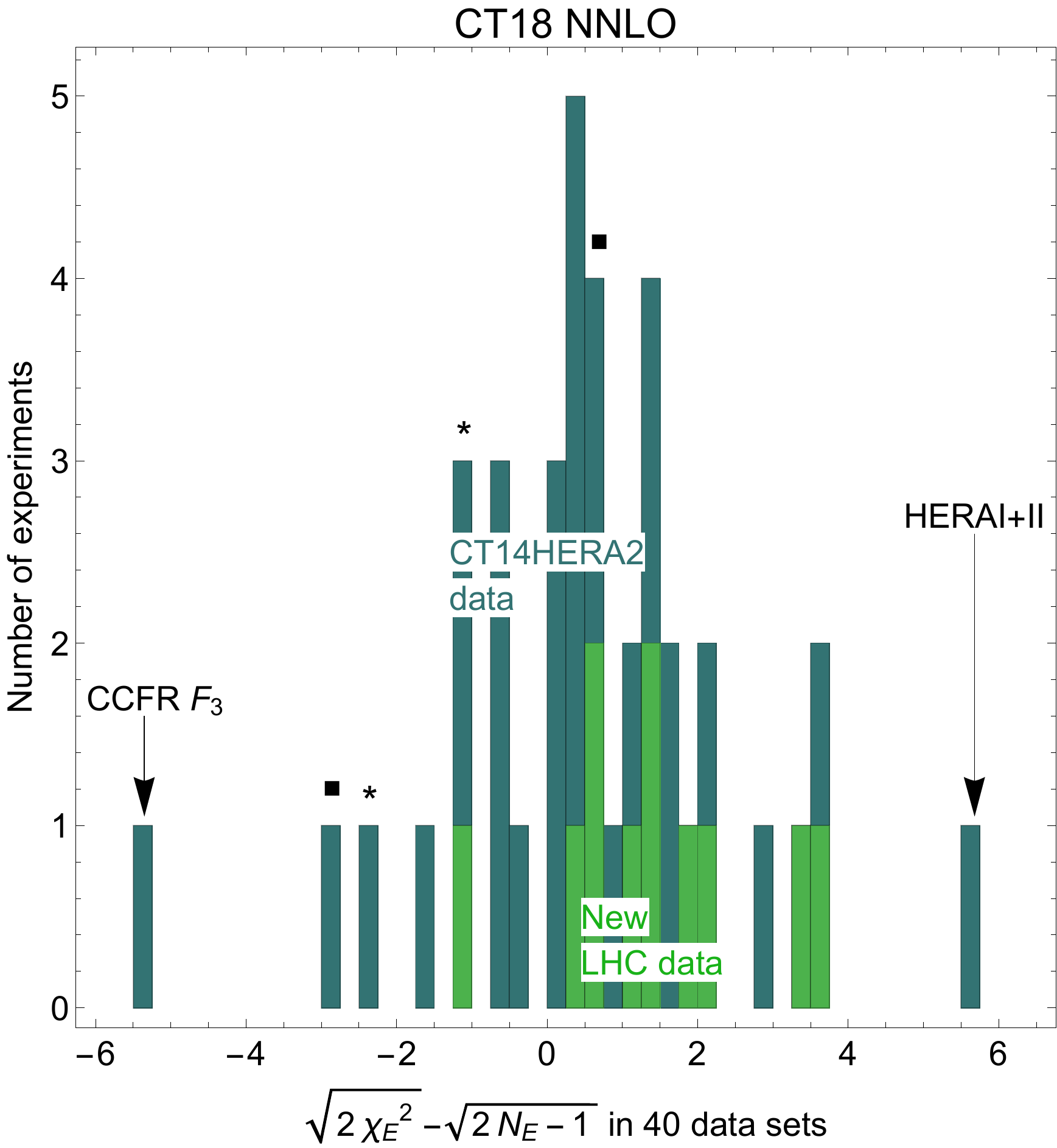}
	\includegraphics[width=0.49\textwidth]{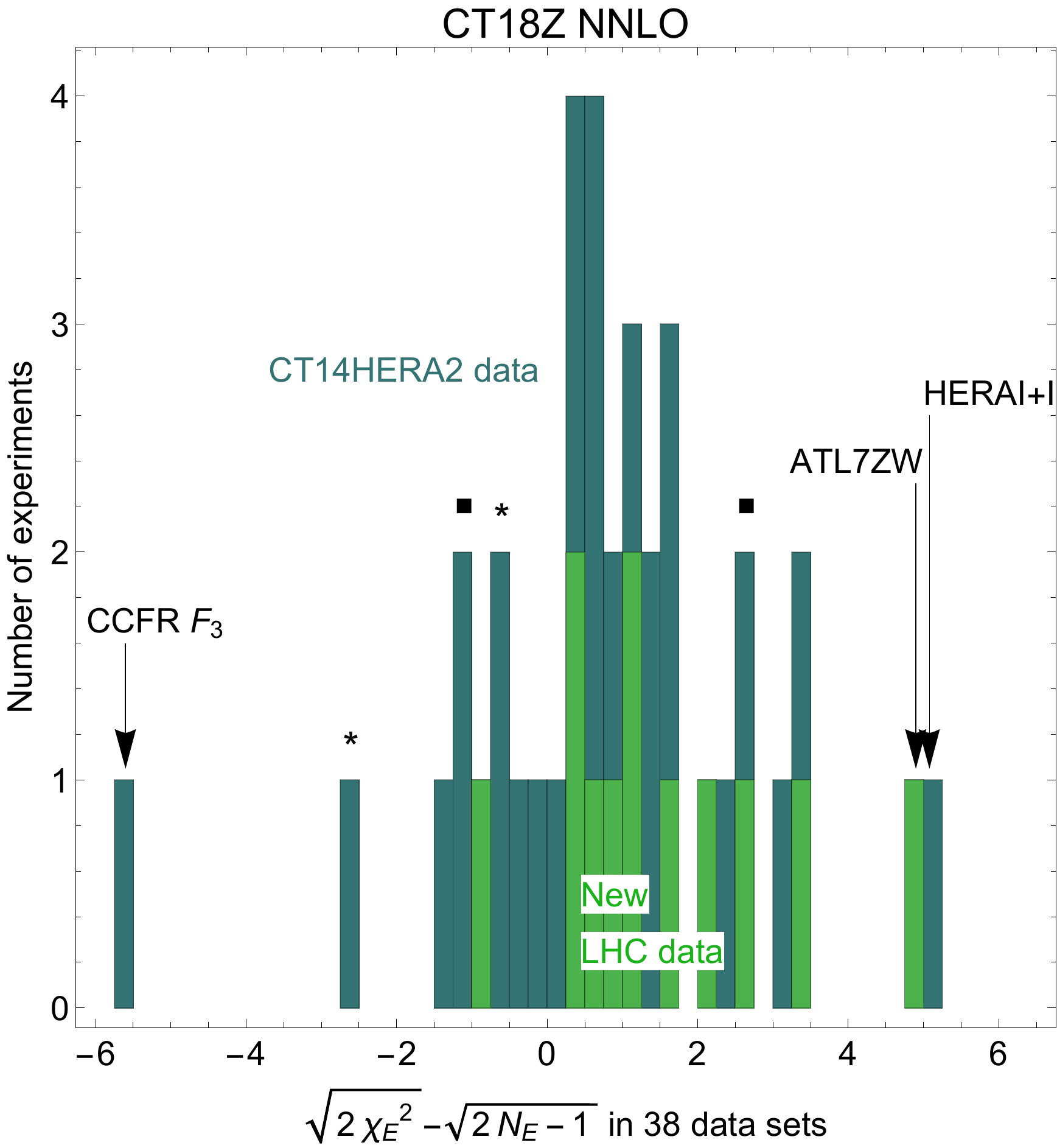}
	\caption{The effective Gaussian variable ($S_n$) distribution of all (a) CT18 data sets, and (b) CT18Z data sets. 
	Two squares and two stars indicate the $S_n$ values for the NuTeV dimuon and CCFR dimuon data, respectively.
\label{fig:sn_ct18_ct18z}}
\end{figure}

\textbf{Selection of new LHC experiments.} Next, we will review the selection
of most promising LHC experiments for the CT18 fit. In this task,
we had to address a recurrent challenge, the presence of statistical
tensions among various (sub)sets of the latest experimental data from
HERA, LHC, and the Tevatron. The quickly improving precision of the collider
data reveals previously irrelevant anomalies either in the experiment
or theory. These anomalies are revealed by applying strong goodness-of-fit
tests \cite{Kovarik:2019xvh}. Figure~\ref{fig:sn_ct18_ct18z}
illustrates the degree of tensions using a representation based on the effective
Gaussian variables $S_E\equiv \sqrt{2\chi_E^2}-\sqrt{2 N_E-1}$ \cite{Lai:2010vv}
constructed from the $\chi^2$ values and numbers of data points for
individual data sets $E$. In a high-quality fit, the probability
distribution for $S_E$ must be approximately a standard normal
distribution (with a unit half-width). In CTEQ-TEA and global
fits from either CTEQ or other groups, 
we in fact observe wider $S_E$ distributions,
cf. Fig.~\ref{fig:sn_ct18_ct18z}, with some most comprehensive
and precise data sets (notably, HERA I+II inclusive DIS
\cite{Abramowicz:2015mha} and ATLAS 7 TeV $Z/W$ production
\cite{Aaboud:2016btc}) having $S_E$ values as high as five units or more.  
The question, then, is how to select the
clean and accurate experiments for the global analysis from the list
that grows day-by-day, while maximally preserving the consistency of
the selected experiments. 

For example, there are many LHC experimental data sets \cite{Rojo:2015acz}
that are potentially sensitive to the PDFs, including novel measurements
in production of high-$p_{T}$ $Z$ bosons, $t\bar{t}$ pairs, heavy
quarks, and $W+c$ pairs. Including all such candidate experiments
into the full global fit is impractical: CPU costs grow quickly with
the number of experimental data sets at NNLO. Poorly fitted experiments
would increase, not decrease, the final PDF uncertainty. The generation
of one error PDF set took several days of CPU time in the CT14 fit
to 33 experiments in a single-thread mode. Adding 20-30 additional
experiments with this setup was thus impossible.

\textbf{Advancements in fitting methodology.} The CTEQ-TEA group resolved
these challenges through a multi-prone effort. We developed two programs
for fast preliminary analysis to identify the eligible experimental
data sets for the global fit. The \texttt{PDFSense} program
\cite{Wang:2018heo} was developed at SMU
to predict quantitatively, and before doing the fit, which data sets
will have an impact on the global PDF fit. The \texttt{ePump} program
\cite{Schmidt:2018hvu} developed at MSU applies PDF reweighting to
quickly estimate the impact of data on the PDFs prior to the global
fit. These programs provide helpful guidelines for the selection of the 
most valuable experiments based entirely on the previously published
Hessian error PDFs.

The CTEQ fitting code was parallelized to allow faster turnaround
time (one fit within few hours instead of many days) on high-performance
computing clusters. For as much
relevant LHC data as possible, we computed our own tables for
APPLGrid/fastNLO fast
interfaces \cite{Kluge:2006xs,Carli:2010rw} for NLO cross sections
(to be multiplied by tabulated point-by-point NNLO/NLO corrections)
for various new LHC processes: production of high-$p_{T}$
Z bosons, jets, $t\bar{t}$ pairs. The APPLgrid tables were cross validated
against similar tables from other groups (available in the public
domain) and optimized for speed and accuracy. 

\textbf{The resulting family of new PDFs} consists of four NNLO PDF
ensembles, and the corresponding NLO ones: the default CT18 ensemble and three
alternative ensemble, designated as CT18A, X, and Z. 
Based on the \texttt{PDFSense} and \texttt{ePump}
studies, eleven new LHC data sets have been included in all four PDF fits,
notably, data at 7 and 8 TeV on lepton pair, jet, and $t\bar{t}$
production. Significant effort was spent on understanding the sources
of PDF uncertainties. Theoretical uncertainties associated with the
scale choice were investigated for the affected processes such as
DIS and high-$p_{T}$ $Z$ production. Other considered theoretical
uncertainties were due to the differences among the NNLO/resummation
codes (FEWZ, ResBos, MCFM, NNLOJet++,...) and Monte-Carlo integration.
The important parametrization uncertainty was investigated by repeating
the fits for 90+ trial functional forms of the PDFs. [Our post-CT10
fits parametrize PDFs using Bernstein polynomials, which simplify
trying a wide range of parametrization forms to quantify/eliminate
potential biases.] As we already mentioned, 
in addition to the default CT18 PDF ensemble,
the other three sets were obtained under alternative assumptions.
(a) The CT18A and CT18Z analyses include high-luminosity ATLAS 7 TeV $W/Z$
rapidity distributions \cite{Aaboud:2016btc} that show some tension
with DIS experiments and prefer a larger strangeness PDF than the
DIS experiments. Inclusion of the ATLAS 7 TeV $W/Z$ data leads to
worse $\chi^2_E$ values (higher $S_E$ values) for dimuon SIDIS
production data sensitive to the strangeness PDF. This can be seen in
the comparison of $S_E$ distributions in Fig.~\ref{fig:sn_ct18_ct18z},
where the $S_E$ values for CCFR and NuTeV dimuon data sets are
elevated in the CT18Z fit on the right, as compared to the CT18 fit on
the left, as a consequence of inclusion of the ATLAS $W/Z$ data in the
CT18Z fit.   
(b) The CT18X and CT18Z fits use an $x$-dependent factorization scale
in NNLO DIS cross sections to mimic enhanced higher-order logarithms
at small Bjorken $x$ and small $Q$. This choice results in the
enhanced gluon PDF at small $x$ and reduced gluon at $x \sim 0.01$, as
discussed above. 
Furthermore, the CDHSW data for DIS on heavy nuclei prefer a
somewhat harder gluon PDF at $x >
0.1$  than other data sets.
In the CT18Z fit, we have removed the CDHSW data.
The combination of these choices in the
CT18Z results in the NNLO Higgs production cross section via gluon
fusion that is
reduced by about 1\% compared to the corresponding
CT14 and CT18 predictions. Thus, the various choices made during the
generation of four CT18(A,X,Z) data sets allow us to more faithfully
explore the full range of the PDF behavior at NNLO that is consistent
with the available hadronic data, with implications for electroweak
precision physics measurements and new physics searches at the LHC. 

\begin{acknowledgments}
	The work of J.~Gao was sponsored by the National Natural Science Foundation
	of China under the Grant No. 11875189 and No.11835005. 
	The work at SMU is supported by the U.S. Department of Energy under Grant No. DE-SC0010129. T.~J.~Hobbs acknowledges support from an EIC Center Fellowship. The work of M. Guzzi is supported by the National Science Foundation under Grant No. PHY1820818. The work of C.-P. Yuan was supported by the U.S. National Science Foundation under Grant No. PHY-1719914, and he is also grateful for the support from the Wu-Ki Tung endowed chair in particle physics. 
	
\end{acknowledgments}


%

\end{document}